\documentclass[twocolumn]{aastex62}
\usepackage{CJK}
\usepackage{longtable}
\usepackage{graphicx}
\usepackage{multirow}
\usepackage{color}
\usepackage{amsmath}
\usepackage{booktabs} 
\usepackage{array}
\usepackage{enumerate}
\usepackage[utf8]{inputenc}

\newcommand{\Ha}{\mbox{H$\alpha$}}      
\newcommand{\Hb}{\mbox{H$\beta$}}       
\newcommand{\Hg}{\mbox{H$\gamma$}}      
\newcommand{\HI}{\mbox{\ion{H}{1}}}     
\newcommand{\HII}{\mbox{\ion{H}{2}}}    
\newcommand{\SII}{\mbox{\ion{S}{2}}}    
\newcommand{\SIII}{\mbox{\ion{S}{3}}}    
\newcommand{\OIII}{\mbox{\ion{O}{3}}}   
\newcommand{\OII}{\mbox{\ion{O}{2}}}    
\newcommand{\NII}{\mbox{\ion{N}{2}}}     

\newcommand{\msun}{\mbox{$M_{\odot}$}}

\newcommand{\Zsun}{\mbox{$Z_{\odot}$}}

\newcommand{\myr}{$M_{\odot}$ yr$^{-1}$}

\begin{document}
\begin{CJK*}{UTF8}{gbsn}

\title{Characterizing Stellar and Gas Properties in NGC 628: Spatial Distributions, Radial Gradients, and Resolved Scaling Relations}

\correspondingauthor{Hu Zou}
\email{zouhu@nao.cas.cn, weipeng@xao.ac.cn}
\author[0000-0002-5674-4223]{Peng Wei (魏鹏)}
\affil{Xinjiang Astronomical Observatory, Chinese Academy of Sciences, Urumqi 830011, China}
\affiliation{National Astronomical Observatories, Chinese Academy of Sciences, Beijing 100101, China}
\author[0000-0002-6684-3997]{Hu Zou (邹虎)}
\affiliation{National Astronomical Observatories, Chinese Academy of Sciences, Beijing 100101, China}
\affiliation{School of Astronomy and Space Science, University of Chinese Academy of Sciences, Beijing 100049, Peopleʼs Republic of China}
\author{Jing Wang}
\affiliation{Kavli Institute for Astronomy and Astrophysics, Peking University, Beijing 100871, Peopleʼs Republic of China}
\author[0000-0002-7660-2273]{Xu Kong (孔旭)}
\affiliation{Key Laboratory for Research in Galaxies and Cosmology, Department of Astronomy, University of Science and Technology of China, Hefei 230026, China}
\author[0000-0001-5066-5682]{Shuguo Ma (马树国)}
\affiliation{National Astronomical Observatories, Chinese Academy of Sciences, Beijing 100101, China}
\affiliation{Xinjiang Astronomical Observatory, Chinese Academy of Sciences, Urumqi 830011, China}
\author{Ruilei Zhou (周瑞蕾)}
\affiliation{National Astronomical Observatories, Chinese Academy of Sciences, Beijing 100101, China}
\affiliation{School of Astronomy and Space Science, University of Chinese Academy of Sciences, Beijing 100049, Peopleʼs Republic of China}
\author{Xu Zhou (周旭)}
\affiliation{National Astronomical Observatories, Chinese Academy of Sciences, Beijing 100101, China}
\author[0000-0003-1845-4900]{Ali Esamdin}
\affiliation{Xinjiang Astronomical Observatory, Chinese Academy of Sciences, Urumqi 830011, China}
\affiliation{School of Astronomy and Space Science, University of Chinese Academy of Sciences, Beijing 100049, Peopleʼs Republic of China}
\author{Jiantao Sun}
\affiliation{Xinjiang Astronomical Observatory, Chinese Academy of Sciences, Urumqi 830011, China}
\affiliation{School of Astronomy and Space Science, University of Chinese Academy of Sciences, Beijing 100049, Peopleʼs Republic of China}
\author{Tuhong Zhong}
\affiliation{Xinjiang Astronomical Observatory, Chinese Academy of Sciences, Urumqi 830011, China}
\author{Fei Dang(党飞)}
\affiliation{Xinjiang Astronomical Observatory, Chinese Academy of Sciences, Urumqi 830011, China}



\begin{abstract}
Building on our previous research of multi-wavelength data from UV to IR, we employ spectroscopic observations of ionized gas, as well as neutral hydrogen gas obtained from the Five-hundred Meter Aperture Spherical Telescope (FAST), to explore the intrinsic processes of star formation and chemical enrichment within NGC 628. Our analysis focuses on several key properties, including gas-phase extinction, star formation rate (SFR) surface density, oxygen abundance, and {\HI} mass surface density. The azimuthal distributions of these parameters in relation to the morphological and kinematic features of FAST {\HI} reveal that NGC 628 is an isolated galaxy that has not undergone recent interactions. We observe a mild radial extinction gradient accompanied by a notable dispersion.  The SFR surface density also shows a gentle radial gradient, characteristic of typical spiral galaxies. Additionally, we find a negative radial metallicity gradient of $-0.44$ dex $R_{25}^{-1}$, supporting the ``inside-out" scenario of galaxy formation. We investigate the resolved Mass-Metallicity Relation (MZR) and the resolved Star Formation Main Sequence (SFMS) alongside their dependencies on the physical properties of both ionized and neutral hydrogen gas. Our findings indicate no secondary dependency of the resolved MZR on SFR surface density or {\HI} mass surface density. Furthermore, we observe that gas-phase extinction and the equivalent width of {\Ha} both increase with SFR surface density in the resolved SFMS.

\end{abstract}



\section{Introduction}\label{sec:intro}

By resolving individual {\HII} regions in nearby spiral galaxies, we can extract a wealth of information at discrete spatial locations through the analysis of emission lines and the underlying stellar continua in high-quality spectra. This includes insights into stellar population properties \citep{sanchezB14a,sanchezB14b,hu18,wei20,parikh21}, mass and luminosity \citep{rosales12,garca19}, absolute or relative gas-phase abundances \citep{kennicutt96,sanchez14,berg15,lin17,hu18,kreckel19,wei20}, and star formation rates \citep{kennicutt12,cat15,gonzalez16,wei20}, 
etc. Moreover, the high spatial resolution achieved through the observation of {\HII} regions in nearby galaxies allows for the differentiation of photoionized regions from other ionizing sources, such as diffuse ionized gas (DIG), supernova remnants, and active galactic nuclei (AGN) \citep{moustakas06,kreckel19}. By aggregating hundreds of high-quality spectra from individual {\HII} regions within a single galaxy, we can conduct a detailed examination of the intricate spatial distribution of stars, dust, and gas. 

While large spectroscopic surveys like the Sloan Digital Sky Survey (SDSS; \citealt{york00}) and the Mapping Nearby Galaxies at APO (MaNGA; \citealt{bundy15}) have facilitated statistical studies of galaxy properties and established global scaling relations, such as the mass-metallicity relation \citep{tremonti04,rosales12,barrera16,sanchez19} and the star formation main sequence \citep{cano19,sanchez19,pessa21}, these often aggregate data across many galaxies. However, understanding how these relations manifest within a single galaxy at high spatial resolution is essential for contextualizing the broader trends observed in larger samples. The resolved mass-metallicity relation (rMZR) and resolved star formation main sequence (rSFMS) shed light on variations in metallicity and star formation across different regions within a galaxy, elucidating the ``inside-out" growth process and localized star formation conditions.

NGC 628, also known as M74 or the Phantom Galaxy, is a grand-design spiral galaxy situated in the constellation Pisces. It has a SFR of 0.15 {\myr} \citep{zaragoza18}, placing it squarely on the main sequence of star formation. Its face-on orientation and isolation from significant external interactions make it an ideal candidate for studying intrinsic physical processes without the confounding effects of recent mergers or strong tidal interactions. This isolation provides a unique opportunity to probe how internal dynamics and gas accretion contribute to star formation and chemical enrichment.

 To analyze the stellar population of NGC 628, we incorporate ultraviolet observations from GALEX \citep{gil07}, 15 intermediate-band data from the Beijing-Arizona-Taipei-Connecticut (BATC) Color Survey of the Sky \citep{fan96}, and infrared observations from the Spitzer Space Telescope \citep{kennicutt03}. These data sets enable us to derive spatially resolved maps of age, metallicity, and reddening. The formation and evolution of different components within NGC 628 are explored in light of these stellar population properties \citep{zou2011c}. As part of a sample of 20 nearby galaxies, NGC 628 has also been studied using long-slit spectroscopy with a 2.16 m telescope at the Xinglong station of the National Astronomical Observatories of China (NAOC) \citep{fan16}. In addition, we employed FAST \citep{nan11} to obtain a deep {\HI} image with high sensitivity and moderate resolution. The integration of single-dish and synthesis data from the THINGS survey \citep{walter08} provides valuable insights into the gas reservoir surrounding NGC 628 and offers potential evidence of tidal interactions with its satellites.
 
 Combining these spectroscopic measurements from {\HII} regions with neutral hydrogen ({\HI}) observations, we derive various physical properties, including extinction, stellar mass surface density (${\Sigma}_{\star}$), SFR surface density ($\Sigma_{\mathrm{SFR}}$), and gas-phase abundance. Analyzing these distributions allows us to investigate the relationships between ${\Sigma}_{\star}$, $\Sigma_{\mathrm{SFR}}$, and gas-phase abundance, as well as their spatial variations across the galaxy. This focused investigation of NGC 628 enhances our comprehension of galaxy evolution by showcasing the role of internal dynamics and progressive gas accretion in shaping isolated spiral galaxies. It offer a more nuanced view of the processes driving star formation and chemical evolution in the local universe.

Basic parameters of NGC 628 utilized in this study are summarized in Table \ref{tab:pars}. This paper is organized as follows: In Section \ref{sec:data}, we provide an overview of the spectroscopic and {\HI} observations, detailing the data reduction processes employed. Section \ref{sec:measurement} outlines the methodology utilized to measure various physical properties of NGC 628. In Section \ref{sec:result}, we present the azimuthal and spatial distributions of these properties, accompanied by relevant comparisons and analyses. Finally, Section \ref{sec:summary} offers a summary of our findings.

\begin{deluxetable}{lcc}[!htb]
\tablecaption{Basic parmeters of NGC 628. \label{tab:pars}}
\tablecolumns{3}
\tablehead{\colhead{Parameter} & \colhead{Value} & \colhead{Reference\tablenotemark{a}}}
\startdata 
          R.A. (J2000)      & $01^{h}36^{m}41.^{s}747$                         & (1) \\
          Decl. (J2000)     & +$15{\arcdeg}47{\arcmin}01.{\arcsec}18$   & (1) \\
          Distance            &  7.3 Mpc                                              & (2) \\
          Inclination          &  7\arcdeg                                             & (2) \\
          Position angle    &  20\arcdeg                                            & (2) \\
 Morphological type       &  SA(s)c                                                  & (1) \\
          Galactic E(B-V)   & 0.062 mag                                              & (3) \\
         $R_{25}$           & 315.{\arcsec}0                                        & (4) \\
         $R_{e}$             & 287.{\arcsec}2                                        & (4) \\
\enddata
\tablenotetext{a}{References. (1) NASA/IPAC Extragalactic Database (NED); (2) \citet{walter08}; (3) \citet{schlafly11}; (4) Third Reference Catalog of Bright Galaxies (RC3) \citep{de95}.}
\end{deluxetable}

\section{Observations and Data Reduction} \label{sec:data}
\subsection{Spectroscopic Data from NAOC 2.16 m Telescope} \label{subsec:obs}

Using the long-slit spectrograph of the 2.16 m telescope and multi-fiber spectrograph of the Multiple Mirror Telescope \citep[MMT;][]{fabricant05}, \citet{kong14} conducted spectroscopic observations of {\HII} regions in 20 nearby large face-on spiral galaxies starting in 2008. These galaxies are too large to be fully encompassed by IFS surveys. With high spatial-resolution spectroscopic data ($<$ 160 pc) from {\HII} regions in M33, M101 and M51, we have examined the distributions of the galaxy properties in details, including dust extinction, metal abundance, star formation rate, and stellar populations \citep{lin17,hu18,wei20}.

The {\HII} regions were selected based on the continuum-subtracted {\Ha} image obtained from the CTIO 1.5 m telescope on October 21, 2001, and sourced from the NASA/IPAC Extragalactic Database (NED)\footnote{\url{https://ned.ipac.caltech.edu}}. The spectroscopic targets were identified as sources with at least 25 pixels (approximately 75 pc in radius) whose {\Ha} fluxes exceed a critical value. Additionally, bright foreground stars from the 2MASS all-sky Point Source Catalog were excluded from the sample.

From 2007 to the end of 2013, a total of 24 nights were allocated for long-slit spectroscopic observations of NGC 628 using the Xinglong 2.16 m telescope. These observations were conducted with the Optomechanics Research Inc. (OMR) Cassegrain spectrograph \citep{fan16}. The 300 line/mm grating was selected, yielding a dispersion of 4.8 $\mbox{\AA}\ \mathrm{pixel}^{-1}$ and a spectral resolution of 10 {\AA}. The wavelength coverage ranges from 3600 to 8000 {\AA} with a blazing wavelength of 5500 {\AA}.
 
\begin{figure}[!htbp]
\center
\resizebox{\width}{!}{
{\includegraphics[width=0.4\textwidth]{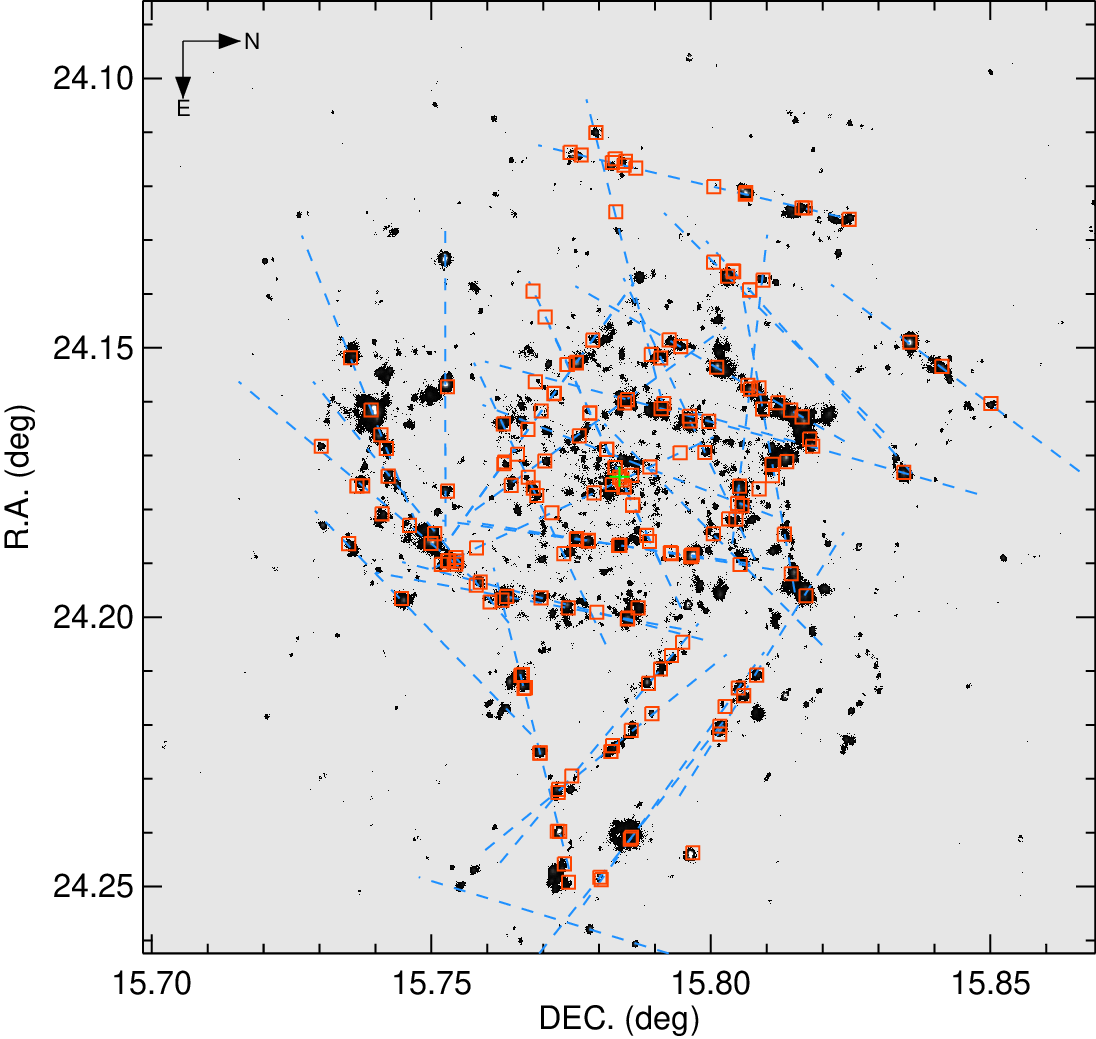}}}
\caption{Map of the slit positions (blue dashed line) used to observe the {\HII} regions in NGC 628. A total of 36 slit positions were observed and 183 spectra were extracted. The background is a gray-scale representation of the continuum-subtracted {\Ha} image. North is oriented to the right and east is oriented down. The center of NGC 628 is marked with a green plus sign, while the red squares indicate the positions from which the spectra were extracted.}
\label{fig:obs}
\end{figure}

Figure \ref{fig:obs} depicts the slit positions overlaid on a continuum-subtracted {\Ha} image of NGC 628. Each slit, measuring 4{\arcmin} in length and 2.\arcsec5 in width (approximately 87.5 pc at the distance of NGC 628) was positioned strategically. At the beginning and end of each observing night, bias and dome flat frames were acquired for the preprocessing of raw CCD images of the science targets, as well as for He-Ar arc lamp and standard stars. Two separate 1800 s exposures were conducted for each target with He-Ar arc lamps used before and after observations for precise wavelength calibration. To subtract the background sky light from the galaxy spectra, we observed the sky background spectra with a 1200 s exposure between the two exposures of each slit position. Spectra of standard stars, selected from the catalog of International Reference Stars (\citealt{corbin91}), were collected for flux calibration. As illustrated in Figure \ref{fig:obs}, a total of 36 slit positions were observed, with two positioned along the major and minor axes of NGC 628. The remaining slits were manually adjusted to maximize coverage of {\HII} regions during observations.

The raw data were processed using the IRAF software\footnote{\url{https://iraf-community.github.io/}}. This included bias subtraction and flat-fielding for the raw data from the scientific targets, standard stars, and He-Ar lamps. Following cosmic-ray removal and sky-background subtraction, the spectrum of each {\HII} region was extracted based on the dispersion trace of the flux standard star. The wavelength calibration was performed using the extracted spectrum of the arc lamp at the corresponding CCD position. Subsequently, the spectra of observed flux standard stars, along with the mean atmospheric extinction coefficients at the Xinglong Station, were used to conduct flux calibration for each wavelength-calibrated spectrum.

In total, 183 spectra were extracted and corrected for Galactic extinction using the extinction law of \citet{car89} and a reddening value of $E(B-V)$ = 0.062 from the foreground Galactic dust map of \citet{schlegel98}. Among these, 127 spectra exhibit either strong stellar continua or prominent emission lines upon visual inspection and were selected for estimating the physical parameters of {\HII} regions. The most distant location of the spectra is located approximately 9.1 kpc from the center of NGC 628. 

\subsection{Atomic Hydrogen Data from FAST}\label{subsec:hiobs}
FAST offers high sensitivity and low side-lobe performance, making it an ideal telescope for mapping extended and low surface-density neutral hydrogen gas. Observations were conducted on November 4, 2022, as part of the FAST Extended Atlas of Selected Targets Survey \citep[FEASTS;][]{wang23}. The data were collected using the Multibeam on-the-fly (OTF) mode with a full width at half maximum (FWHM) of the raw beam of $\sim$2.9{\arcmin} at a frequency of 1.42 GHz \citep{jiang20}. The observed sky region covered a Right Ascension range of 23.45{\arcdeg} $<$ $\alpha$ $<$ 24.90{\arcdeg} and a Declination range of 15.14{\arcdeg} $<$ $\delta$ $<$ 16.48{\arcdeg} along with a velocity scope of 492.7-780.9 km s$^{-1}$, utilizing the 19-beam receiver system of FAST \citep{jiang19,jiang20}. This receiver operates in dual polarization mode and covers a frequency band from 1050 to 1450 MHz. During the observations, a 10 K noise diode was intermittently activated for 1 second every 60 seconds to ensure consistent calibration. The backend instrument employed a Spec(W+N) spectrometer, which has a bandwidth of 500 MHz and 65,536 channels, providing a spectral resolution of 1.67 km s$^{-1}$ for {\HI} 21 cm line observations. 

Raw data reduction was performed following a standard pipeline developed for reducing radio single-dish image data, with detailed procedures outlined in \citet{wang23}. Four primary processes were employed: radio frequency interference (RFI) flagging, calibration, imaging, and baseline flattening. After Hanning smoothing, the velocity resolution was degraded from 1.67 km s$^{-1}$ to 4.88 km s$^{-1}$ through a threshold-based smooth and clipping source-finding algorithm. The final data cube was generated in the standard FITS format with a grid spacing of 0.5{\arcmin} in the image plane.

\begin{figure*}[!ht]
\center
\resizebox{\width}{!}{
{\includegraphics[width=0.8\textwidth]{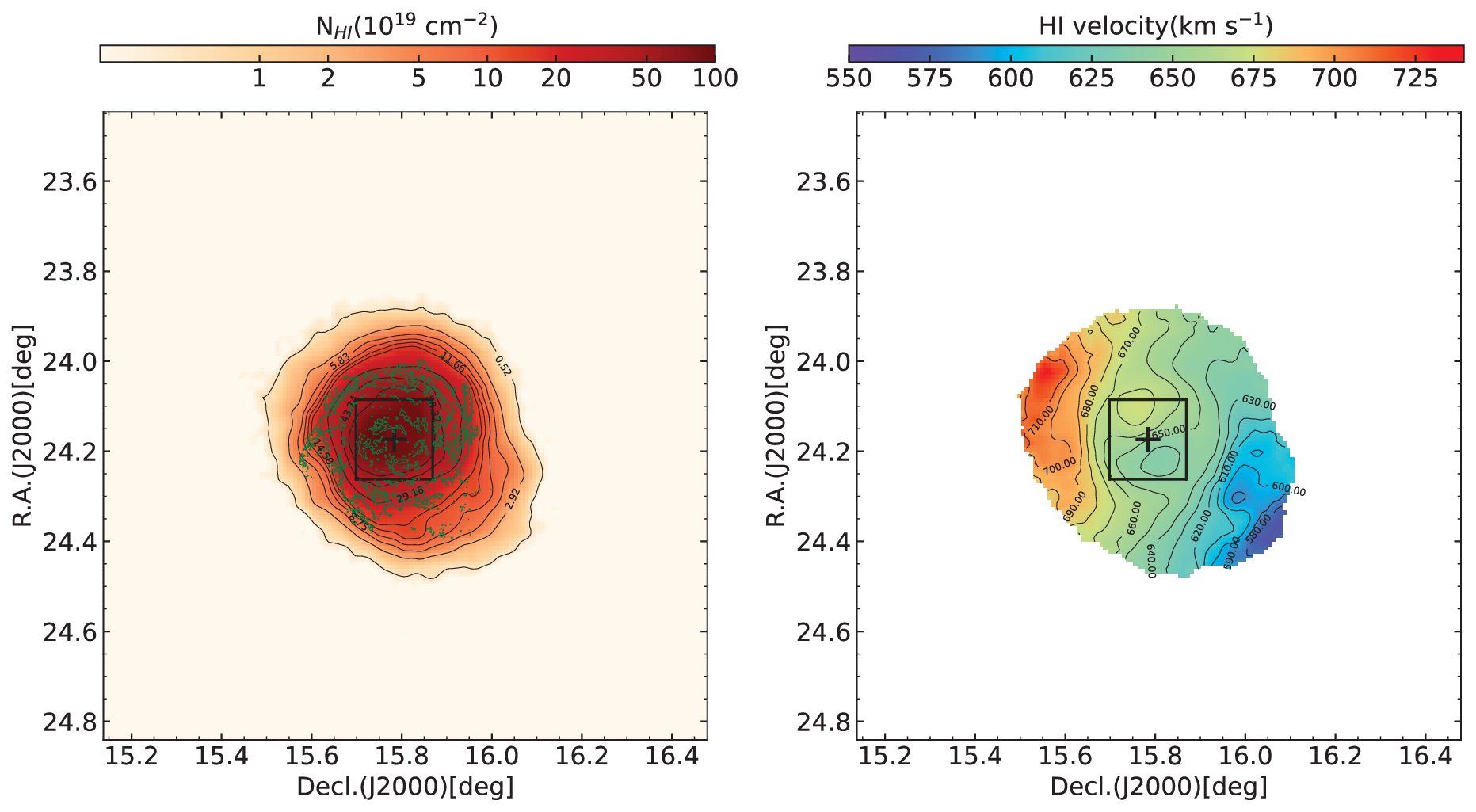}}}
\caption{The {\HI} integrated intensity map and the velocity field of NGC 628 from the FAST {\HI} observations. There is a significantly broader coverage of {\HI} column density map compared to THINGS. Left: the black contours represent the FAST {\HI} column density integrated over 492-780 km s$^{-1}$, with contour levels begin at 0.18 Jy km s$^{-1}$, which corresponds to a column density of 5.2 $\times$ 10$^{18}$ cm$^{-2}$. The green contours represent the {\HI} intensity from the THINGS survey, integrated over 588-735 km s$^{-1}$. The black box indicates the sky region shown in Figure \ref{fig:obs}. The center of NGC 628 is marked with plus sign. Right: the velocity contours begin at 580 km s$^{-1}$ and increase in steps of 10 km s$^{-1}$.}
\label{fig:hi}
\end{figure*}

The left panel of Figure \ref{fig:hi} presents the {\HI} column-density map of NGC 628, extracted from the moment 0 image. This map is overlaid with {\HI} integrated flux density contours obtained from the Very Large Array (VLA) as part of the THINGS \citep{walter08}. The FAST {\HI} map demonstrates significantly broader coverage of the galaxy compared to THINGS due to its high sensitivity. From this map, we derive a total integrated flux of 569 Jy km s$^{-1}$, and an {\HI} mass of 7.1$\times$10$^{9}${\msun} for NGC 628, which is 1.86 times larger than that from the THINGS measurement (302 Jy km s$^{-1}$ and 3.8$\times$10$^{9}${\msun}).  This indicates a substantial amount of diffuse {\HI} gas outside the {\HI} disk, which was previously unresolved in the THINGS map. The {\HI} coverage where the column density $N_{\HI} \ge 10^{19}$cm$^{-2}$ extends $\sim$90 kpc across. The moment 1 image highlights rotational velocity gradients in the disk regions of NGC 628, as shown in the right panel of Figure \ref{fig:hi}. The {\HI} velocity field spans a range of 492-780 km s$^{-1}$ and exhibits a gradient decreasing from southwest to northeast. The central velocity of the {\HI} profile is 656.1 km s$^{-1}$, consistent with the THINGS measurement of 659.1 km s$^{-1}$.

\section{Analyzing Stellar and Ionized Gas Properties} \label{sec:measurement}

\subsection{Full spectrum fitting and emission-line intensity measurements} \label{subsec:specfit}
\begin{figure*}[!htbp]
\center
\resizebox{\width}{!}{
{\includegraphics[width=0.9\textwidth]{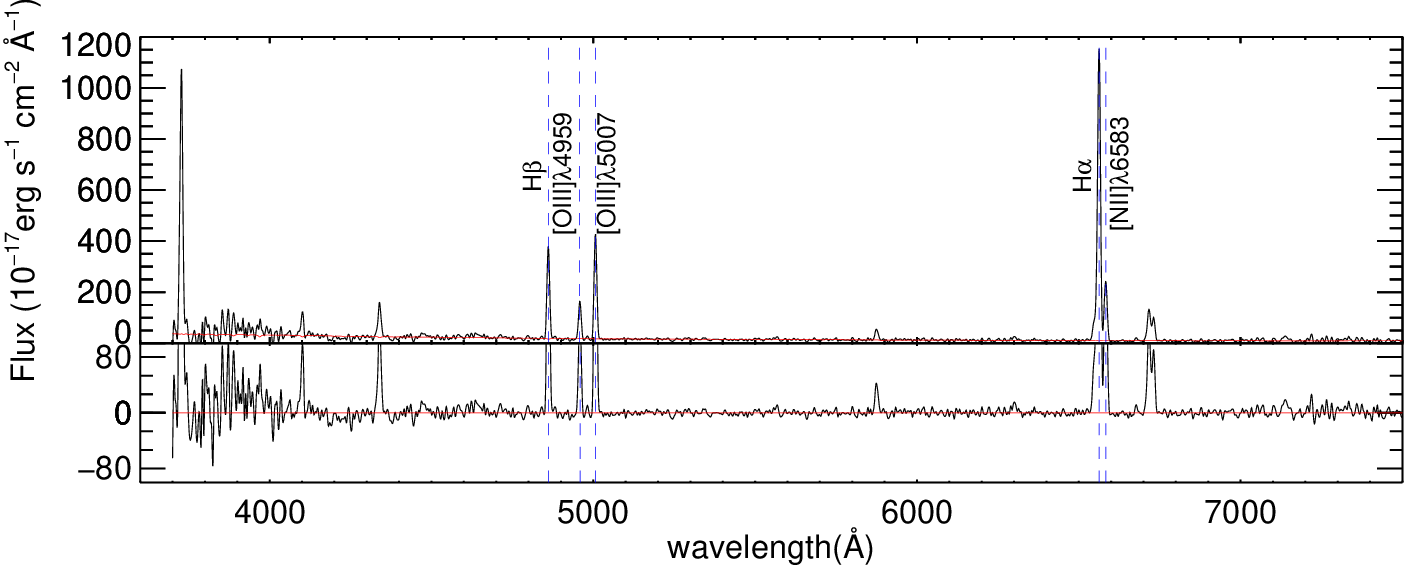}}}
\caption{A high signal-to-noise ratio (S/N) spectrum of an {\HII} region observed with the Xinglong 2.16 m telescope (black) is shown in the upper panel. The best-fit model spectrum is overplotted in red. In the lower panel, the residual spectrum, representing the difference between the observed and model spectra, is displayed. The horizontal red line indicates zero residual. The vertical dashed lines mark the emission lines utilized in this study.}
\label{fig:spectra}
\end{figure*}

We resampled all spectra with a wavelength step of 1 {\AA} in the range of 3700--7500 {\AA} using IRAF’s \textit{dispcor} procedure, in accordance with the recommendations of the STARLIGHT\footnote{\url{http://www.starlight.ufsc.br}} spectral synthesis code \citep{cid05}. To separate the emission lines from the underlying stellar continuum, we utilize STARLIGHT to model the spectra with templates from \citet{bru03}\footnote{\url{http://www.bruzual.org}}. The observed spectrum was treated as a linear combination of simple stellar population (SSP) templates. Employing a \citet{chab03} initial mass function (IMF), we fitted each spectrum using a grid of 150 SSP templates, enabling us to extract the underlying stellar continuum and estimate the dust absorption. The templates spanned a stellar age range of 1 Myr to 18 Gyr with 25 different ages, and include six metallicities ranging from 0.005 to 2.5 Z$_{\sun}$. We applied the extinction law of \citet{car89} to correct for dust extinction effects. Additionally, optical nebular emission lines, Wolf-Rayet features, and atmospheric absorption lines were masked out using the standard emission-line masks provided by STARLIGHT. An example of the spectral modeling is illustrated in Figure \ref{fig:spectra}, which also presents the best-fit model spectra.

After subtracting the stellar continuum, we measure the strengths of emission lines using Gaussian fitting. During this process, the flux ratios of [{\NII}]$\lambda$6548 to [{\NII}]$\lambda$6583 and [{\OIII}]$\lambda$4959 to [{\OIII}]$\lambda$5007 were fixed at 1/3. The standard errors for the line strengths are derived from the combination of continuum errors and those introduced by the Gaussian fitting. The continuum error is calculated as the root mean square (rms) noise within a 200 {\AA} spectral region centered on the emission line, excluding areas affected by nebular emission.

Given the low spectral resolution and signal-to-noise ratio (S/N) of the collected spectra, temperature-sensitive auroral lines or doublets (e.g., [{\OIII}]$\lambda$4363, [{\NII}]$\lambda$5755, [{\SIII}]$\lambda$6312, and [{\OII}]$\lambda\lambda$7320, 7330) are not reliably identifiable in most {\HII} regions. Consequently, we focus our further analysis on the strong emission lines of {\Hb}, [{\OIII}]$\lambda$5007, {\Ha}, and [{\NII}]$\lambda$6583.

\subsection{Redenning estimation and extinciton corrections} \label{subsec:reddencor}
The relative intensities of the Balmer lines are nearly independent of the physical conditions of the gas, such as electronic density and temperature. Therefore, the de-reddening of spectra can be derived using Balmer line ratios, such as {\Ha}/{\Hb}, {\Ha}/{\Hg}, and {\Hb}/{\Hg}. Assuming a temperature of $T_{e}$ = 10$^{4}$ K and a density of $n_{e}$ =10$^{2}$ cm$^{-3}$, the reddening value is determined from {\Ha}/{\Hb} using the following formula:
\begin{small}
\begin{equation}
E(B-V)=\frac{1}{-0.4 \times (K_{\mathrm{H\alpha}}-K_{\mathrm{H\beta}})}\log_{10}{\frac{(\mathrm{H\alpha}/\mathrm{H\beta})_{\mathrm{obs}}}{(\mathrm{H\alpha}/\mathrm{H\beta})_{\mathrm{int}}}}.
\end{equation}
\end{small}
Here, ({\Ha}/{\Hb})$_\mathrm{int}$ is the intrinsic flux ratio of 2.86 for case B recombination. ({\Ha}/{\Hb})$_\mathrm{obs}$ is the observed flux ratio. $K_\mathrm{\Ha}$ and $K_\mathrm{\Hb}$ represent the values of the reddening law at the wavelengths of the {\Ha} and {\Hb} lines, respectively. When $E(B-V) < 0$ or the observed {\Ha}/{\Hb} $\le$ 2.86, we set E(B-V) = 0, although {\Ha}/{\Hb} $\le$ 2.86 can also occur physically in {\HII} regions. 

We compare the differences in E(B-V) using the extinction law of \citet{car89} and the attenuation law of \citet{calzetti00}. The difference is within approximately 0.03 mag, which is smaller than the typical measurement error of around 0.1 mag \citep{wei20}. Therefore, the extinction curves have little effect on this attenuation correction, a conclusion that is also supported by previous literature \citep[e.g.,][]{cat15}. The gas-phase extinction in the $V$ band is calculated by $A_{V} = 3.1 E(B-V)$. Based on the attenuation law of \citet{car89} (where $R_{V}$ = 3.1, $K_\mathrm{\Ha}$ = 2.53 and $K_\mathrm{\Hb}$ = 3.61), all the emission line intensities are corrected for the gas-phase extinction.

\subsection{BPT diagrams to exclude objects affected by AGN} \label{sec:bpt}
\begin{figure}[!htb]
\center
\resizebox{\width}{!}{
{\includegraphics[width=0.48\textwidth]{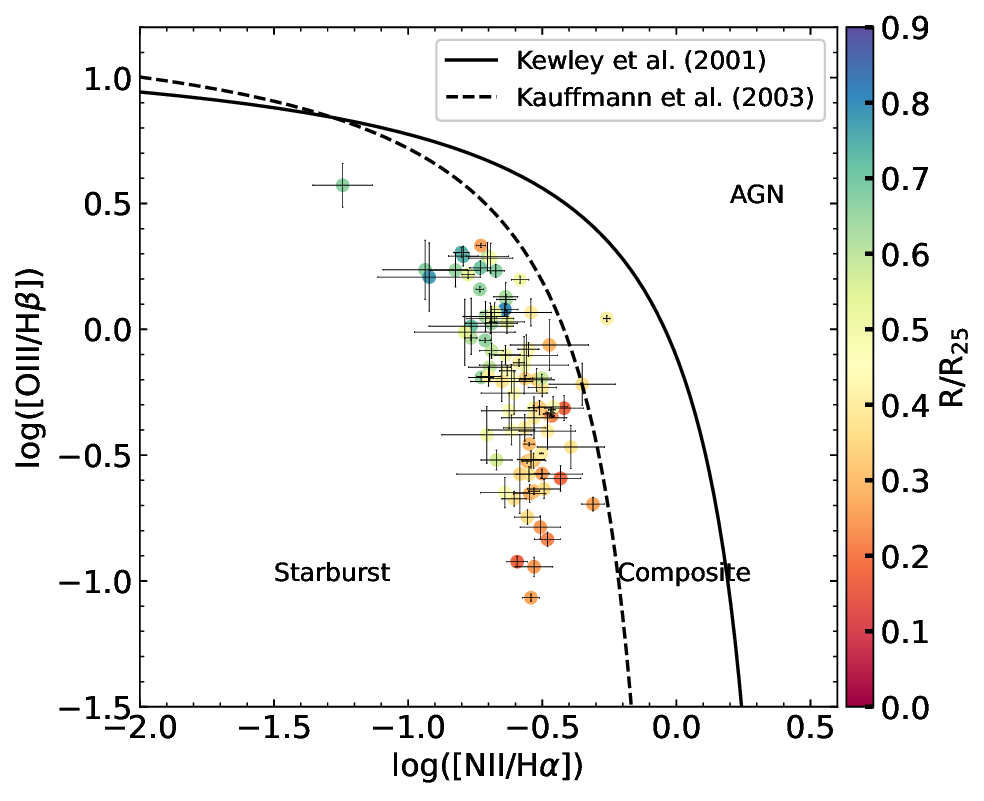}}}
\caption{The resolved diagnostic BPT diagram displaying [{\NII}]$\lambda$6583/{\Ha} versus [{\OIII}]$\lambda$5007/{\Hb}. Each point on the diagram are color-coded according to radial position. The solid and dashed lines denote the demarcation curves established by \citet{kewley01} and \citet{kauffmann03}, respectively.}
\label{fig:bpt}
\end{figure}

In Figure \ref{fig:bpt}, the Baldwin-Phillips-Terlevich (BPT) diagram displays the ratio of the [{\OIII}]$\lambda$5007 to {\Hb}  versus [{\NII}]$\lambda$6583 to {\Ha} \citep{baldwin81}, providing insights into the excitation properties of the spectra. The demarcations from \citet{kewley01} and \citet{kauffmann03} categorize the spectra into three distinct regions: star-forming regions, regions excited by shocks or active galactic nuclei (AGN), and composite regions. Figure \ref{fig:bpt} further illustrates color-coded radial positions, indicating that the inner regions exhibit higher [{\NII}]$\lambda$6583 and lower [{\OIII}]$\lambda$5007 compared to the outer regions. This pattern implies a negative metallicity gradient, with detailed analysis provided in Section \ref{subsubsec:metallicity}.

Figure \ref{fig:bpt} presents 85 spectral samples on the BPT diagram, all of which correspond to star-forming regions situated close to or below the Kauffmann demarcation curve with only one exceptions. Additional spectra, not included in the BPT diagram due to insufficient S/Ns for emission lines, are located farther from the nucleus of NGC 628. Consequently, these areas are classified as star-forming regions.

\subsection{SFR surface density} \label{subsec:SFRmethod}
Hydrogen recombination lines are sensitive to the shortest timescales of star formation, allowing for the instantaneous SFR estimates. The \citet{chab03} IMF yields total stellar masses comparable to those from the \citet{kroupa03} IMF when normalized to high stellar masses. Thus, \citet{chomiuk11} indicated that SFRs derived from the \citet{kroupa03} IMF are applicable to those from the \citet{chab03} IMF. Assuming a \citet{kroupa03} IMF with solar metallicity, SFRs can be calculated from the extinction-corrected {\Ha} luminosity ($L({\Ha}_\mathrm{corr})$) using the equation from \citet{hao11}:
\begin{equation}\label{eq:sfr}
\mathrm{SFR}\ (\mathrm{M_{\sun}\ yr^{-1}}) = 5.4\times10^{-42}L({\Ha}_\mathrm{corr})(\mathrm{erg}\ \mathrm{s^{-1}}).
\end{equation}

The SFR surface density ($\Sigma_{\mathrm{SFR}}$) is calculated as the SFR per unit area measured within rectangle apertures, where the length ($l$) represents the aperture size in arcsec used for extracting the spectrum and the width ($w$) corresponds to the slit width in arcsec. The $\Sigma_{\mathrm{SFR}}$ is computed using the following expression: 
\begin{equation}
\Sigma_{\mathrm{SFR}}\ (\mathrm{M_{\sun}\ yr^{-1}\ pc^{-2}}) = \frac{\mathrm{SFR}}{l \times w \times (\frac{d}{206265})^{2}},
\end{equation}
where $d$ is the distance to the galaxy in pc.

\subsection{Gas-phase oxygen abundance} \label{subsec:O/Hmethod}
The determination of gas-phase metallicity via the direct $T_{e}$ method requires the calculation of the electron temperature, which is based on the ratio of faint auroral line intensities, such as [{\OIII}]$\lambda$4363 \citep{osterbrock06}. However, auroral lines are often too faint to be detected in the majority of our spectra. Consequently, we use stronger emission lines to calculate the metallicity, including  [{\OII}]$\lambda$3727, {\Hb}, [{\OIII}]$\lambda$5007, {\Ha}, [{\NII}]$\lambda$6583, and [{\SII}]$\lambda\lambda$6717, 6731. Based on the relative intensity of these emission lines, a variety of strong-line indicators have been proposed to estimate the oxygen abundance. These indicators rely on the relationship between line ratios and the oxygen abundance derived from the direct $T_{e}$ method \citep[e.g., O3N2, N2,][]{pp04,marino13}, or from photoionization models \citep[e.g., R23, N2O2,][]{pilyugin05,kewley02}. Since only a limited number of spectra contain detections of the [{\OII}]$\lambda$3727 line, the metallicity estimators associated with this line are not applicable. Thus, we adopt the empirical calibrator based on the O3N2 ratio:
\begin{equation}\label{eq:o3n2}
\mathrm{O3N2}=\log_{10}{(\frac{[\OIII]\lambda5007/\Hb}{[\NII]\lambda6583/\Ha})}.
\end{equation}
This ratio is weakly affected by differential dust attenuation and exhibits a monotonic, single-valued behavior within its applicable range.

We use the improved O3N2 empirical oxygen calibration from \citet{marino13} to calculate the oxygen abundance, expressed as
\begin{equation}
12+\log{(\mathrm{O/H})}=8.533-0.214\times \mathrm{O3N2}.
\end{equation}
This indicator is valid for the range of -1.1 $<$ O3N2 $<$ 1.7, which corresponds to oxygen abundances of 12+log(O/H) $>$ 8.0 dex. The calibration uncertainty is reported to be 0.18 dex \citep{marino13}. All our spectra yield O3N2 ratios within this valid range. The random error in the metallicity is estimated by propagating the errors of the emission line measurements.

\subsection{Stellar mass surface density} \label{subsec:mass}
The stellar mass $M_{\star}$ is inferred from modeling the multi-wavelength photometric SEDs. Multi-wavelength images of NGC 628, ranging from UV to IR (including data from GALEX, XMM-OM, BATC, 2MASS, and Spitzer), are obtained from \citet{zou2011c}. The stellar population models and SED fitting methodology are described in \citet{wei21}. A total of 23 bands are used to extract the multi-band SED at the respective slit position of each {\HII} region. Photometric fluxes were obtained via rectangular aperture photometry, carried out using the Astropy package in \textit{PHOTUTILS} \citep{bradley22}. The rectangular apertures are defined with dimensions of $l \times w$, and are roated to match the position angle of the slit relative to the major axis of the galaxy. Utilizing the \citet{chab03} IMF, Padova 1994 evolutionary tracks, and a delayed-exponential SFH, we construct a library of 10$^{5}$ model spectra using the \citet{bru03} code, which are randomly sampled across a range of parameter spaces of age (1 Myr to 13.5 Gyr) and metallicity (0.2 to 2.0 \Zsun). 

The stellar masss surface density ${\Sigma}_{\star}$ is calculated as
\begin{equation}
\Sigma_{\mathrm{\star}}\ (\mathrm{M_{\sun}\ \ pc^{-2}}) = \frac{M_\mathrm{\star}}{l \times w \times (\frac{d}{206265})^{2}},
\end{equation}
where $M_{\star}$ is derived from SED fitting, and $d$ is the distance to the galaxy in pc.

\section{Results and Discussion} \label{sec:result}
In Table \ref{tab:flux}, we present the measurements of emission lines and their corresponding spectral properties. A total of 127 spectra exhibit {\Ha} and {\Hb} fluxes with a S/N greater than 5. Additionally, the S/Ns for [\OIII]$\lambda$5007 and [\NII]$\lambda$6583 are also required to exceed 5. Consequently, valid flux measurements for these emission lines were obtained from 85 spectra. In Table \ref{tab:flux}, the {\Hb} flux represents the absolute line strength, while the fluxes for the other lines are relative, normalized to the {\Hb} flux.

\begin{deluxetable*}{lcccccccccccccc}
\tablecaption{Emission-line measurements and spectral properties\label{tab:flux}}
\tablewidth{3800pt}
\tabletypesize{\scriptsize}
\tablehead{\colhead{ID}  & \colhead{R.A.} & \colhead{Decl.}  & \colhead{$R/R_{25}$} & \colhead{[{\OIII}]} & \colhead{{\Ha}} & \colhead{[{\NII}]} & \colhead{{\Hb}} & \colhead{EW({\Ha})} & \colhead{$E(B-V)$} &  \colhead{12+log(O/H)} & 
\colhead{$\log\Sigma_{\mathrm{SFR}}$} & \colhead{$\log\Sigma_{\star}$} & \colhead{{\HI} flux} & \colhead{$\log\Sigma_{\HI}$}\\
\colhead{(1)}  & \colhead{(2)}  & \colhead{(3)}  & \colhead{(4)} & \colhead{(5)} & \colhead{(6)} & \colhead{(7)} & \colhead{(8)} & \colhead{(9)} & \colhead{(10)} & \colhead{(11)} & \colhead{(12)} & \colhead{(13)} & \colhead{(14)}& \colhead{(15)}} 
\startdata
  1 &      01:36:45.341 &  15:45:15.803  &    0.374   &             &     2.759   &     0.830 &     303.353 &   324.728 &           &          &  -0.818  &   7.510  &   0.513  &   5.856 \\
    &                   &                &            &             &     0.015   &     0.016 &       8.833 &    13.767 &           &          &   0.002  &          &          &         \\
  2 &      01:36:45.492 &  15:45:11.124  &    0.391   &     0.587   &     2.870   &     0.907 &     375.547 &   575.251 &    0.051  &   8.476  &  -0.436  &   7.530  &   0.491  &   5.837 \\
    &                   &                &            &     0.049   &     0.169   &     0.054 &      32.887 &    27.650 &    0.025  &   0.014  &   0.002  &          &          &         \\
  3 &      01:36:45.643 &   15:45:6.444  &    0.407   &     1.104   &     2.551   &     1.404 &     281.706 &   606.511 &           &   8.468  &  -0.723  &   7.490  &   0.461  &   5.809 \\
    &                   &                &            &     0.015   &     0.025   &     0.036 &       6.725 &    83.909 &           &   0.004  &   0.004  &          &          &         \\
  4 &      01:36:39.391 &  15:45:46.404  &    0.261   &     0.202   &     2.882   &     1.409 &    1035.881 &   554.874 &    0.121  &   8.615  &  -0.824  &   7.780  &   0.598  &   5.622 \\
    &                   &                &            &     0.015   &     0.127   &     0.063 &      66.825 &    22.417 &    0.019  &   0.011  &   0.003  &          &          &         \\
  5 &      01:36:41.767 &   15:46:2.389  &    0.187   &             &     1.744   &     0.952 &     117.558 &   330.837 &           &          &  -1.261  &   8.220  &   0.273  &   5.583 \\
    &                   &                &            &             &     0.049   &     0.047 &      10.654 &    56.020 &           &          &   0.012  &          &          &         \\
  6 &      01:36:42.252 &   15:46:5.664  &    0.178   &     0.486   &     2.900   &     1.106 &     564.294 &   345.304 &    0.216  &   8.511  &  -0.790  &   8.130  &   0.267  &   5.572 \\
    &                   &                &            &     0.057   &     0.228   &     0.092 &      65.369 &    24.540 &    0.033  &   0.019  &   0.005  &          &          &         \\
  \dots &  \dots  &  \dots  &    \dots   &     \dots   &     \dots   &     \dots &     \dots &   \dots &    \dots  &    \dots      &  \dots  &   \dots  &   \dots  &   \dots \\
  \dots &  \dots  &  \dots  &    \dots   &     \dots   &     \dots   &     \dots &     \dots &   \dots &    \dots  &    \dots      &  \dots  &   \dots  &   \dots  &   \dots \\
120 &      01:36:41.042 &  15:46:13.045  &    0.156   &             &     2.952   &     0.930 &     992.422 &   998.959 &    0.489  &          &  -0.327  &   8.320  &   0.490  &   5.614 \\
    &                   &                &            &             &     1.045   &     0.329 &     524.114 &   187.720 &    0.151  &          &   0.005  &          &          &         \\
121 &      01:36:39.922 &  15:46:35.293  &    0.117   &             &     3.034   &     0.807 &   11060.521 &   209.147 &    0.911  &          &  -0.092  &   8.360  &   0.892  &   5.460 \\
    &                   &                &            &             &     1.102   &     0.293 &    5996.469 &     7.077 &    0.155  &          &   0.004  &          &          &         \\
122 &      01:36:38.450 &   15:47:4.523  &    0.152   &             &     2.927   &     0.903 &    3386.817 &   119.887 &    0.359  &          &  -0.956  &   8.220  &   0.863  &   5.559 \\
    &                   &                &            &             &     0.329   &     0.102 &     567.092 &     1.650 &    0.048  &          &   0.003  &          &          &         \\
123 &      01:36:44.722 &   15:44:6.720  &    0.572   &     1.105   &     2.885   &     0.584 &    1393.031 &   628.204 &    0.136  &   8.376  &  -1.139  &   7.030  &   1.885  &   6.053 \\
    &                   &                &            &     0.151   &     0.286   &     0.058 &     205.715 &    28.343 &    0.042  &   0.023  &   0.002  &          &          &         \\
124 &      01:36:47.165 &  15:44:40.777  &    0.513   &     1.166   &     2.883   &     0.606 &    1932.103 &   496.683 &    0.126  &   8.374  &  -1.191  &   7.210  &   1.085  &   5.814 \\
    &                   &                &            &     0.100   &     0.177   &     0.038 &     176.744 &    14.156 &    0.026  &   0.014  &   0.002  &          &          &         \\
125 &      01:36:40.385 &  15:43:49.080  &    0.613   &     0.924   &     2.563   &     0.440 &     756.181 &  1484.878 &           &   8.377  &  -1.452  &   6.850  &   0.886  &   5.872 \\
    &                   &                &            &     0.024   &     0.016   &     0.011 &      33.602 &   201.964 &           &   0.005  &   0.003  &          &          &         \\
126 &      01:36:42.156 &  15:44:12.082  &    0.538   &     0.974   &     2.917   &     0.474 &    1299.133 &  1189.455 &    0.307  &   8.365  &   0.098  &   7.380  &   1.009  &   5.849 \\
    &                   &                &            &     0.286   &     0.617   &     0.102 &     410.164 &   197.277 &    0.091  &   0.052  &   0.004  &          &          &         \\
127 &      01:36:43.395 &  15:44:28.212  &    0.492   &     1.575   &     2.777   &     0.726 &     478.033 &  2078.813 &           &   8.366  &  -1.200  &   7.390  &   1.949  &   6.068 \\
    &                   &                &            &     0.034   &     0.028   &     0.022 &      33.632 &   700.923 &           &   0.007  &   0.004  &          &          &         \\
\enddata
\tablecomments{(1): object number. (2)-(3): equatorial coordinate (J2000). (4): scaled galactocentric distance, where $R$ is the galactocentric distance and $R_{25}$ is the apparent major isophotal radius. (5)-(7): relative fluxes of [{\OIII}]$\lambda$5007,[\NII]$\lambda$6583 and {\Ha} that are normalized to the {\Hb} flux.  (8): {\Hb} flux in unit of $10^{-17}\mathrm{erg~s^{-1}~cm^{-2}}$. (9): equivalent width of {\Ha} line in {\AA} . (10): gas-phase reddening in mag. (11): metallicity in dex. (12): SFR density in $\rm M_{\sun} yr^{-1} kpc^{-2}$. (13): logarithmic stellar mass surface density in $\rm M_{\sun} kpc^{-2}$. Each row is followed by another row presenting the error. (14): THINGS {\HI} flux in unit of Jy km s$^{-1}$. (15): logarithmic THINGS {\HI} mass surface density in unit of $\rm M_{\sun} kpc^{-2}$.}
\end{deluxetable*}

\subsection{Spatial distributions and gradients of physical properties}
\subsubsection{Azimuthal distribution for any evidence of tidal interaction} \label{subsubsec:azimuthal}
\begin{figure*}[!htbp]
\center
\resizebox{\width}{!}{
{\includegraphics[width=0.9\textwidth]{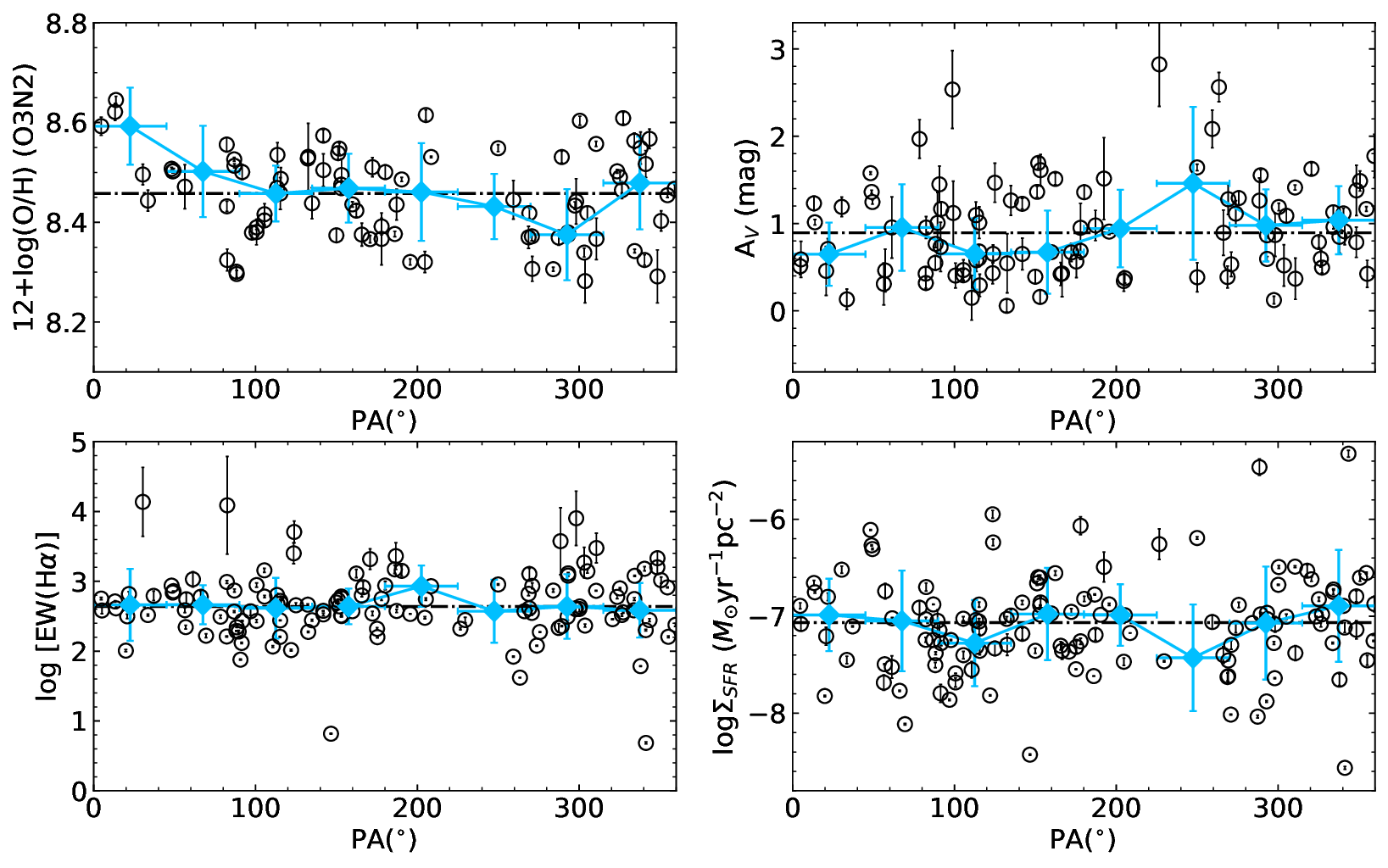}}}

\caption{Azimuthal distributions of oxygen abundance (top left), dust extinction (top right), EW({\Ha}) (bottom left) and $\Sigma_{\mathrm{SFR}}$ (bottom right) for the {\HII} regions in NGC 628. These four physical parameters exhibit minimal variations in the azimuthal distribution. The position angles are deprojected angles on the disk of NGC 628, measured counterclockwise from the north towards the east. In each panel, the blue filled diamonds with vertical bars indicate the median values binned according to PA. The dot-dashed line represents the median value across all data points.}
\label{fig:metalaz}
\end{figure*}

In Figure \ref{fig:metalaz}, we present the azimuthal distributions of oxygen abundance, dust extinction, EW({\Ha}), and $\Sigma_{\mathrm{SFR}}$ for {\HII} regions in NGC 628. The sample of {\HII} regions has been categorized into eight distinct groups based on their projected position angles (PAs) to compute the median values within each bin. The azimuthal oxygen distribution is depicted in the upper left panel of Figure \ref{fig:metalaz}. Notably, there is a significant local fluctuation in oxygen abundance at PAs between $0^{\circ}$ and $90^{\circ}$. We attribute this sudden increase in abundance to the absence of {\HII} region samples located beyond 5 kpc. Additionally, a subtle decrease in abundance is observed at PAs around $270^{\circ}$ to $315^{\circ}$, where the majority of {\HII} regions in this PA bin are situated far from the center of NGC 628. Consequently, these local variations are related to the sparse disbribution of {\HII} samples in certain areas.

The azimuthal extinction distribution is presented in the upper right panel of Figure \ref{fig:metalaz}. The gas-phase extinction in $A_{V}$ is primarily distributed within the range of 0.1 to 1.5 mag, with an overall median value of 0.89 mag. For the median values of $A_{V}$ binned by PA, the differences between the overall median and the binned medians are less than 0.25 mag, with standard deviations of about 0.4 mag. The sixth PA bin ($\sim225^{\circ}-270^{\circ}$) is an exception, exhibiting a deviation of 0.57 mag due to a limited number of samples in that region. This indicates that the gas-phase extinction in NGC 628 displays a nearly uniform azimuthal distribution, despite the presence of local variations. 

The azimuthal distributions of EW({\Ha}) and $\Sigma_{\mathrm{SFR}}$ are shown in the bottom row of Figure \ref{fig:metalaz}, from left to right. The plots reveal minimal variations in the azimuthal distribution of these two physical parameters. Overall, our analysis suggests that the physical properties of NGC 628 exhibit a nearly uniform azimuthal distribution, indicating that NGC 628 is a relatively isolated galaxy with limited interactions with neighboring galaxies. In contrast, for NGC 5194, it has been observed that extinction in the northern spiral arms is lower than in the southern arms, likely due to its proximity and interactions with NGC 5195 \citep{wei20}.

High-sensitivity {\HI} imaging is essential for uncovering structures with low column densities. The left panel of Figure \ref{fig:hi} illustrates a significant presence of low column density {\HI} gas both within and surrounding NGC 628, as detected in the high-sensitivity images obtained from FAST. However, we find no evidence linking NGC 628 to its satellite galaxies at the column density level of 5.2$\times$10$^{18}$cm$^{-2}$(3$\sigma$). Furthermore, the extended structure lacks any visible tidal tails extending in other directions. The right panel of Figure \ref{fig:hi} demonstrates that the {\HI} velocity field within the disk region is relatively regular and undisturbed. These morphological and kinematic characteristics suggest that NGC 628 is an isolated galaxy with minimal evidence of recent interactions in its evolutionary history. 

The black box in Figure \ref{fig:hi} delineates the boundary of the stellar disk of NGC 628. Beyond this boundary, the {\HI} column density in the inner disk remains relatively flat, whereas the extended outer disk exhibits significant warping and a steep decline in {\HI} column density compared to the optical disk. This distinction between the inner and outer disks is also reflected in the kinematic map. These observations suggest that the inner disk (up to the stellar disk boundary) and the warped outer {\HI}-disk are distinct components, potentially formed during different epochs \citep{kruit07}. The inner disk likely formed initially, either through a rapid monolithic process or a more gradual hierarchical process, while the warped outer disk formed later from the infall of gas with higher angular momentum from a different orientation. Furthermore, \citet{wang11} proposed that as {\HI} gas becomes enriched, the specific star formation rate (sSFR) in the outer regions increases relative to the inner regions, indicating that the outer regions of {\HI}-rich galaxies are younger. Consequently, the accreted external gas triggers star formation within the disk, resulting in outward growth alongside the disk and illustrating an ``inside-out" galaxy growth scenario for NGC 628.

\subsubsection{Gas-phase extinction distribution} \label{subsec:dust}
\begin{figure*}[!ht]
\center
\resizebox{\width}{!}{
{\includegraphics[width=0.8\textwidth]{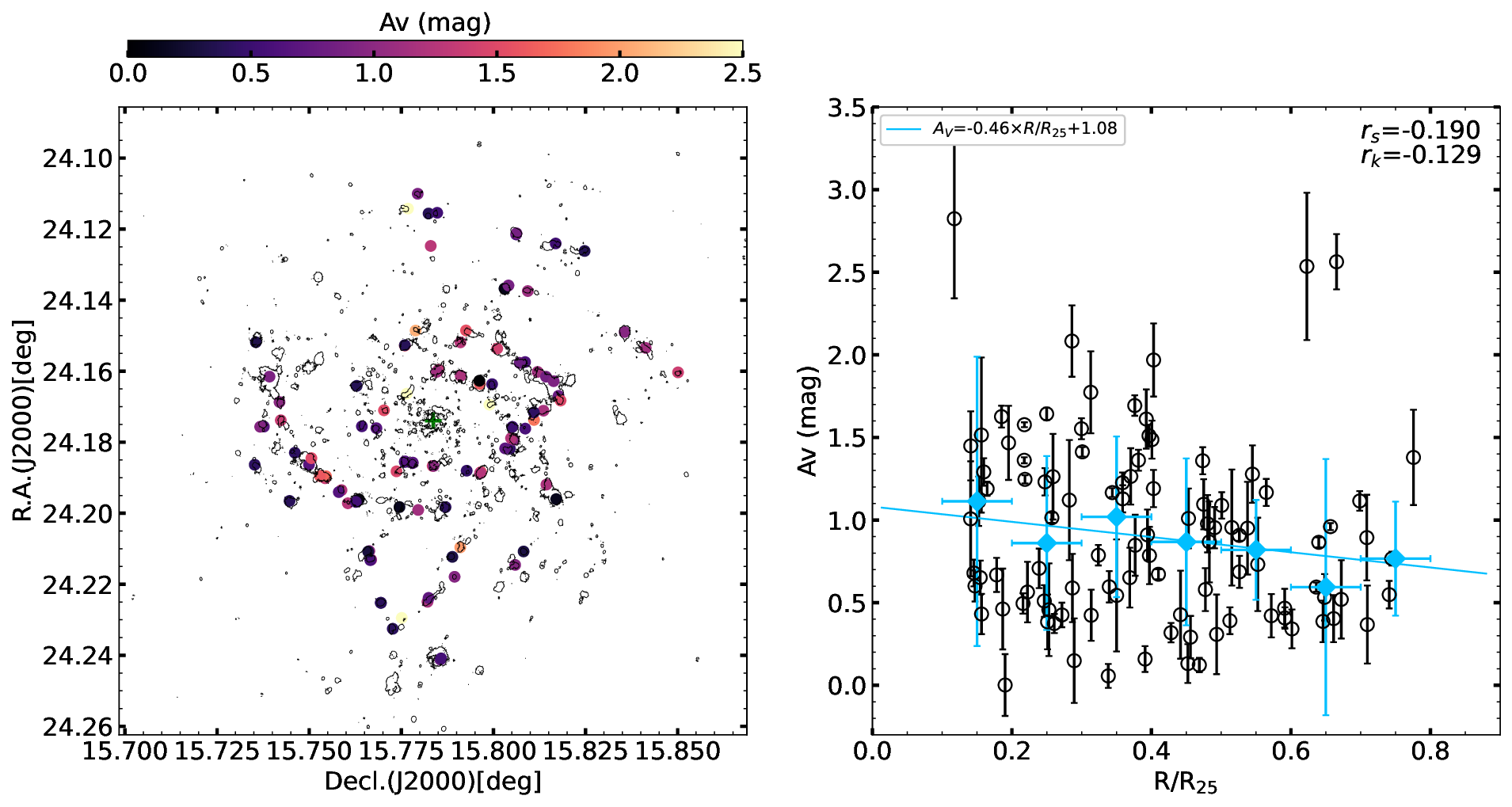}}}
\caption{Two-dimensional distribution and radial distribution of the gas-phase extinction in NGC 628. Left: the contours represents the isophotal shapes of the {\Ha} emission, with the center of NGC 628 indicated by a plus sign. Right: the blue diamonds with vertical error bars represent the median extinction and its standard deviation in each radial bin, while the horizontal bars indicate the range of each bin. The blue line shows the result of the linear fit.}
\label{fig:av}
\end{figure*}

The two-dimensional gas-phase extinction distribution of NGC 628 are illustrated in the left panel of Figure \ref{fig:av}. The map shows no significant variations in the azimuthal distribution of gas-phase extinction; instead, local variations appear to be more pronounced. This indicates that local physical conditions are primary factors influencing the dust properties or gas-phase extinction in NGC 628. Similar findings have been corroborated by analysis of the largest sample of galaxies (approximately 10,000 targets) observed in the SDSS-IV MaNGA survey \citep{barrera22}.

Based on the spatially resolved observations from the MaNGA survey, it is clear that extinction patterns exhibiting radial trends are common among late-type galaxies. Notably, the gradients of these trends demonstrate minimal variation across a range of stellar masses \citep{barrera22}. The right panel of Figure \ref{fig:av} illustrates the deprojected radial distribution of gas-phase extinction $A_{V}$, revealing a relatively large dispersion in radial extinction values. The Spearman rank correlation coefficient ($r_{s}$) and Kendall correlation coefficient ($r_{k}$) presented in the figure suggest a mild radial extinction gradient in NGC 628. Both coefficients are non-parametric statistics assessing the strength and direction of associations based on concordances and discordances in paired observations, with values ranging from -1 to +1. Positive values indicate a positive relationship, negative values signify a negative relationship, and a value of 0 indicates no association. While $r_{s}$ is suited for smaller samples and weaker correlations, $r_{k}$ is more robust against outliers and provides greater precision with strong correlations in small datasets.

To further analyze the gradient, we divide the extinctions inferred from the Balmer decrement into seven distinct bins, based on their galactocentric distance, normalized by the galactic radius R$_{25}$. For each radial bin, we calculate the median extinctions and their standard deviations. Additionally, we perform a linear fit to these median extinctions and yielding a fitted radial gradient of -0.46$\pm$0.17 mag R$_{25}^{-1}$. 

Furthermore, we utilize multi-wavelength imaging data to estimate the level of stellar extinction, revealing a gentle gradient in the radial distribution of reddening across the galaxy \citep{zou2011c}. The concurrent analysis of the radial gradients of both gas-phase and stellar extinctions indicates a higher concentration of dust and gas in the inner regions of NGC 628 compared to its outer areas, suggesting a radial-dependent accumulation of these interstellar components.

\subsubsection{Gas-phase oxygen abundance distribution} \label{subsubsec:metallicity}
\begin{figure*}[!htb]
\center
\resizebox{\width}{!}{
{\includegraphics[width=0.8\textwidth]{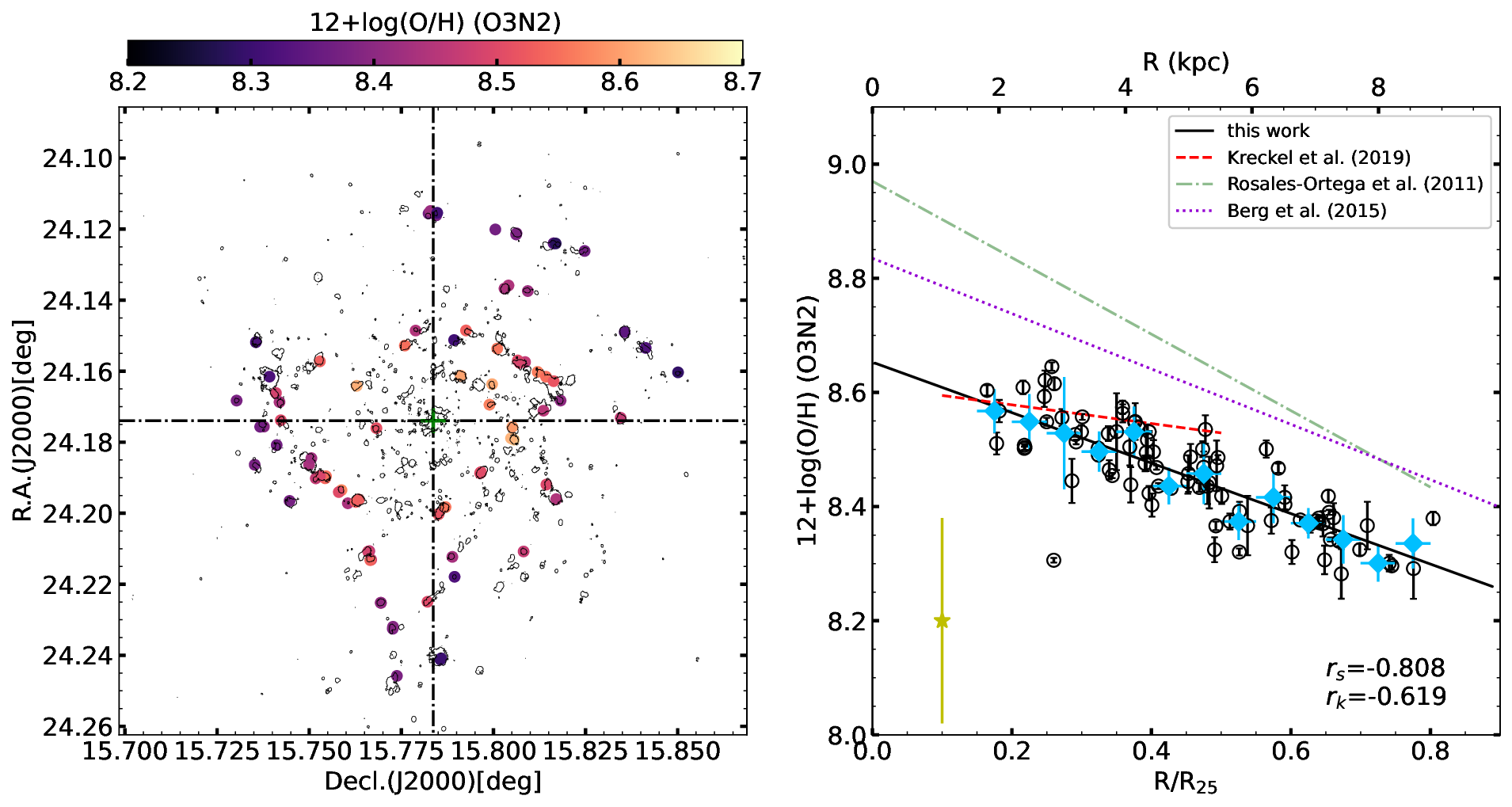}}}
\caption{Two-dimensional distribution and deprojected radial distribution of the oxygen abundance of the {\HII} regions in NGC 628. It exhibits a negative metallicity gradient across the discs. Left: the contours display the isophotal shapes of the {\Ha} emission with the center of NGC 628 marked by a plus sign. Right: the black open circles represent the derived oxygen abundance based on the O3N2 calibrator, with a negative slope shown in this study at $0.15 < R/R_{25} < 0.9$. The blue diamonds indicate the median values of each radial bin, with horizontal bars showing the range of each bin. The green dot-dashed line represents the negative gradient by \citet{rosales11} using the O3N2 calibrator of \citet{pp04}. The red dashed line shows the slope derived from \citet{kreckel19} with MUSE data using the O3N2 calibrator of \citet{marino13} for $0.1 < R/R_{25} < 0.5$. The violet dotted line presents the negative slope measuring a direct abundance with \citet{berg15}. The yellow star indicates the system uncertainty of the O3N2 calibrator of \citet{marino13}.}
\label{fig:o3n2}
\end{figure*}

A total of 85 {\HII} regions in NGC 628 have the gas-phase oxygen abundances estimated using the O3N2 calibration. It is important to note that different strong-line calibrators can yield varying absolute metallicity values. Figure \ref{fig:o3n2} presents the spatial map and radial distribution of oxygen abundance in NGC 628. As shown in the left panel, the central regions of the galaxy present higher oxygen abundances compared to its outer regions, indicating a clear radial gradient. The right panel displays the radial profile of the oxygen abundance. We perform a linear fit to the relationship between oxygen abundances and deprojected galactocentric distances. For the range of $0.15 < R/R_{25} < 0.9$, the linear relation can be expressed as
\begin{eqnarray}
&12&+\log{(\mathrm{O/H})}[\mathrm{O3N2}] \nonumber \\
&=& (8.653\pm0.018) + (-0.404\pm0.034) \times R\left(\mathrm{dex}\ R_{e}^{-1}\right) \nonumber \\
&=& (8.653\pm0.018) + (-0.443\pm0.037) \times R\left(\mathrm{dex}\ R_{25}^{-1}\right)\nonumber \\
&=& (8.653\pm0.018) + (-0.040\pm0.004) \times R\left(\mathrm{dex}\ \mathrm{kpc}^{-1}\right). \label{equ:grad1}
\end{eqnarray}

In previous studies, the metallicity gradient estimates for NGC 628 were determined using the temperature-sensitive auroral lines or various strong-line calibration methods with data from IFS surveys or multi-object spectrographs. The results obtained from different diagnostic techniques are illustrated in the right panel of Figure \ref{fig:o3n2}. Our measured radial negative gradient of oxygen abundance closely aligns with the abundance gradient (-0.485 $\pm$ 0.122 dex R$_{25}^{-1}$) reported by \citet{berg15}, who analyzed temperature-sensitive auroral lines from 45 individual {\HII} regions. The lower metallicity (y-intercept) observed in the radial relation compared to \citet{berg15} is likely due to methodological and calibration differences. The O3N2 strong-line method used here may systematically underestimate metallicity, especially at higher abundances, compared to the direct $T_{e}$ method \citep{marino13}. Additionally, \citet{berg15} note that the [{\OIII}]$\lambda$4363 auroral line, essential for direct $T_{e}$ measurements, is weak in high-metallicity regions, leading to measurement uncertainties and potentially higher y-intercepts in their metallicity gradients. Using the empirical O3N2 calibrator, \citet{rosales11} and \citet{kreckel19} calculated abundance gradients based on IFS data from PINGS and MUSE surveys, respectively. Despite employing the same O3N2 ratio, \citet{rosales11} derived a steeper abundance gradient of approximately -0.66 $\pm$ 0.07 dex R$_{25}^{-1}$ through the O3N2 empirical oxygen calibration methods of \citet{pp04}, contrasting with our results obtained using the calibration methods of \citet{marino13}. The negative slope (-0.164 $\pm$ 0.033 dex R$_{25}^{-1}$) derived with MUSE data in the range of $0.1 < R/R_{25} < 0.5$ is shallower than our findings.

In support of an ``inside-out'' galaxy growth scenario, various observations have confirmed the presence of negative radial gradients in the gas-phase metallicity across the discs of nearby galaxies. NGC 628, in particular, exhibits a negative metallicity gradient. Notably, extensive surveys, such as PINGS, CALIFA, MaNGA and MUSE have provided substantial datasets of galaxies in the local universe, enabling meaningful statistical analyses on abundance gradients and their relationships with morphology, stellar mass, and environment density \citep{sanchezM18, sanchez19}. Figure 15 in \citet{sanchez19} presents radial profiles of oxygen abundance for galaxies of varying morphology and stellar mass. For galaxies with the stellar masses between $10^{9.5}$ M$_{\sun}$ and $10^{10.5}$ M$_{\sun}$ of Sc type within the range of $R< R_e$, the typical gradient is about -0.06 dex $R_e^{-1}$. Notably, NGC 628 presents a significantly steeper metallicity gradient (-0.404 $\pm$ 0.034 dex R$_{e}^{-1}$) compared to galaxies with similar stellar mass and morphological type. Additionally, a radial stellar metallicity gradient of -0.05 dex $\mathrm{kpc}^{-1}$ was derivied for the region of $2.5 < R < 6.3\ \mathrm{kpc}$ \citep{zou2011c}, which closely aligns with the radial O/H abundance gradient at $2.0 < R < 8.5\ \mathrm{kpc}$ (-0.040 $\pm$ 0.004 dex $\mathrm{kpc}^{-1}$).

The central bulge of NGC 628 is identified as a pseudobulge, formed gradually through internal secular processes. This is supported by its nearly exponential luminosity profile, ongoing star formation, and the presence of nuclear spiral structures \citep{zou2011c}. These characteristics indicate that both the disk and the pseudobulge in NGC 628 have evolved via secular mechanisms. The non-axisymmetric spiral arm configuration likely drives gas inflow from the galaxy outskirts to its center, fueling star formation and promoting pseudobulge growth. As an isolated galaxy, external triggers for star formation in NGC 628 are less possible. Instead, accretion of cold gas from the intergalactic medium is a significant factor in initiating such activity. The inward transport of gas triggers star formation, which rapidly enhances the metallicity of the pristine gas within NGC 628, resulting in a steeper metallicity gradient.

\subsubsection{EW({\Ha}) and $\Sigma_{\mathrm{SFR}}$ Distributions} \label{subsubsec:esfr}
\begin{figure*}[!htb]
\center
\resizebox{\width}{!}{
{\includegraphics[width=0.8\textwidth]{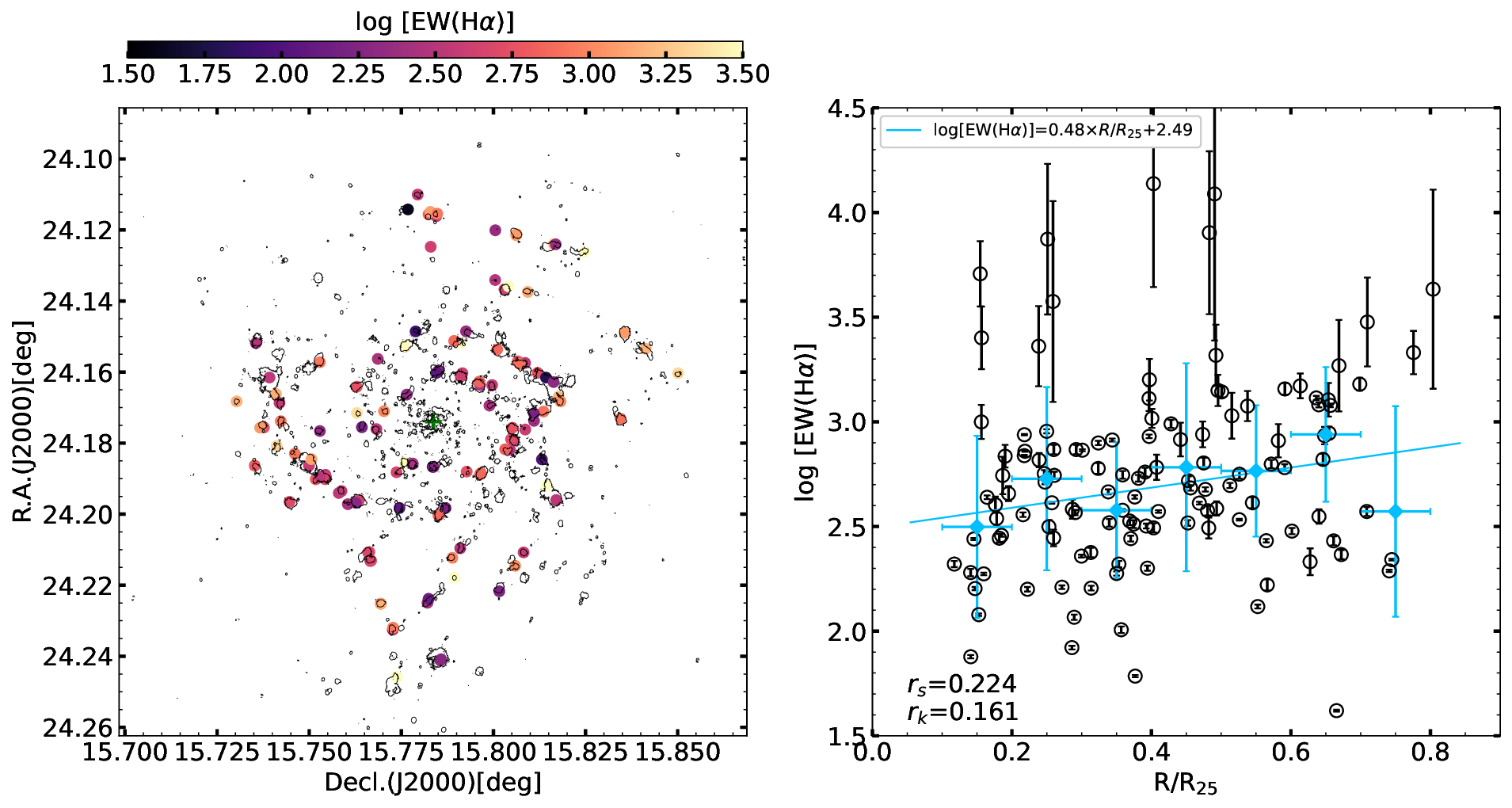}}}
\caption{Two-dimensional distribution and deprojected radial distribution of EW({\Ha}) in logarithmic scale in NGC 628. Left: the contours represent the isophotal shapes of the {\Ha} emission, with the center of NGC 628 marked by a plus sign. Right: the blue diamonds with vertical error bars indicate the median EW({\Ha}) and its standard deviation for each radial bin, while the horizontal bars show the range of each bin. The blue line illustrates the gradient of the median EW({\Ha}) across the radial bins.}
\label{fig:ewha}
\end{figure*}

The equivalent width of the {\Ha} emission line is widely used as a tracer of star formation activity in galaxies. Figure \ref{fig:ewha} illustrates the two-dimensional and deprojected radial distributions of EW({\Ha}) in NGC 628. Consistent with the trends observed in the predominant population of late-type galaxies as reported by \citet{barrera22}, NGC 628 also exhibits a positive gradient in EW({\Ha}). As noted in \citet{kong04}, the EW({\Ha}) is sensitive to the ratio of present to past SFRs. The positive gradient of EW({\Ha}) in NGC 628 suggests that the star formation began earlier in the bulge and inner disk regions, which harbor an older stellar population compared to the outer disk. Consequently, the stellar ages in the bulge and inner disk are older than those in the outer disk. In fact, \citet{zou2011c} report that the bulge and inner disk of NGC 628 have undergone an evolutionary history of about 7 Gyr, while a significant influx of gas into the outer disk likely triggered star formation around 2--3 Gyr ago.

\begin{figure*}[!htb]
\center
\resizebox{\width}{!}{
{\includegraphics[width=0.8\textwidth]{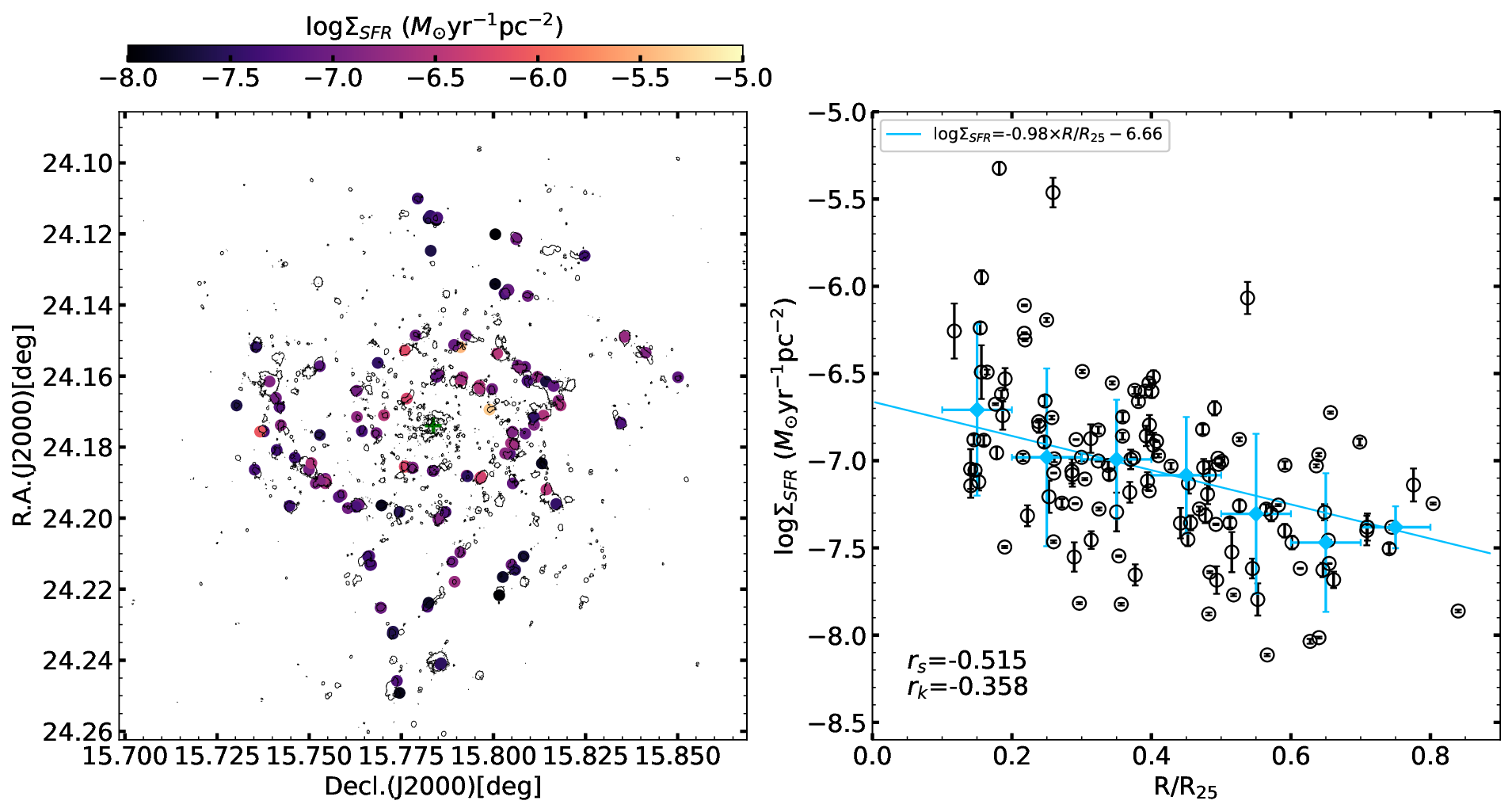}}}
\caption{Two-dimensional distribution and deprojected radial distribution of the SFR surface density in logarithmic scale in NGC 628. It exhibits a steady decline with galactocentric distance, albeit with notable scatter. Left: the contours represent the isophotal shapes of the {\Ha} emission with the center of NGC 628 marked by a plus sign. Right: the blue diamonds with vertical error bars represent the median $\Sigma_{\mathrm{SFR}}$ and its standard deviation for each radial bin, while the horizontal bars indicate the range within each bin. The blue line shows the linear fit to the obtained median $\Sigma_{\mathrm{SFR}}$ across the radial distribution.}
\label{fig:esfr}
\end{figure*}

The left panel of Figure \ref{fig:esfr} presents the distribution of the $\Sigma_{\mathrm{SFR}}$ across the {\HII} regions in NGC 628. Notably, there are no significant variations in the azimuthal distribution of $\Sigma_{\mathrm{SFR}}$. The inner disk exhibits a higher $\Sigma_{\mathrm{SFR}}$ compared to the outer disk. Furthermore, \citet{sanchez11} derived the $\Sigma_{\mathrm{SFR}}$ distribution using the PINGS IFS data, with values ranging from $10^{-6}$ to $10^{-8}$ {\myr}$\mathrm{pc^{-2}}$, which aligns closely with our findings. 

The monotonic decrease in gas-phase oxygen supports an ``inside-out" scenario, where increasing gas infall timescales with increasing distance lead to SFR variations and chemical gradients. This scenario predicts a monotonic decrease in SFR with galactocentric distance. We also present the deprojected radial distribution of $\Sigma_{\mathrm{SFR}}$ in the right panel of Figure \ref{fig:esfr}. This radial distribution in NGC 628 shows a steady decline with increasing galactocentric distance, which aligns with the suggested ``inside-out" growth scenario, although the radial profile exhibits a relatively large dispersion. A linear fit to the radial distribution yields a gradient of -0.98$\pm$0.14 dex R$_{25}^{-1}$. Several studies utilizing the IFS data \citep[e.g.][]{gonzalez16,sanchez19,barrera22} have investigated the radial variations of $\Sigma_{\mathrm{SFR}}$ in relation to the stellar mass ($\mathrm{M_{\star}}$) and the morphological types of galaxies in the local universe. These studies consistently find that late-type galaxies, regardless of their stellar mass, exhibit similar gradients in $\Sigma_{\mathrm{SFR}}$ \citep{barrera22}. 

Morphologically classified as an Sc galaxy, NGC 628 has an esitmated total stellar mass of about 1.2 $\times$ $10^{10}$ {\msun} \citep{zaragoza18}. Compared to the gradients derived by \citet{barrera22} for galaxies of the same morphological type and similar stellar mass, our slope is steeper. The spiral structure of NGC 628 generates a nonaxisymmetric potential that facilitates the inward migration of gas from the outskirts towards its core, leading to a concentrated gas reservoir that enhances star formation activity. \citet{zou2011c} utilized a range of photometric data from UV to IR wavelengths to confirm the presence of a young circumnuclear ring within the bulge, indicative of an ongoing vigorous star formation episode.

\subsection{Spatially resolved scaling relations of NGC 628}\label{subsec:scale}
\subsubsection{Relationship between stellar mass surface density and metallicity} \label{subsubsec:rmzr}
 
The mass-metallicity relation links the primary outcomes of stellar evolution. This connection has been firmly established over a broad range of galaxy masses on a global galactic scales \citep{tremonti04,zhao10}. Furthermore the shape of this relation remains consistent regardless of whether single aperture or spatially resolved spectroscopic data is utilized. On spatial scales of kiloparsec and sub-kiloparsec, studies have also revealed the presence of a resolved MZR in the local universe \citep{rosales12,moran12,gao18,erroz19,sanchez19}. 

\begin{figure}[!ht]
\center
\resizebox{\width}{!}{
{\includegraphics[width=0.42\textwidth]{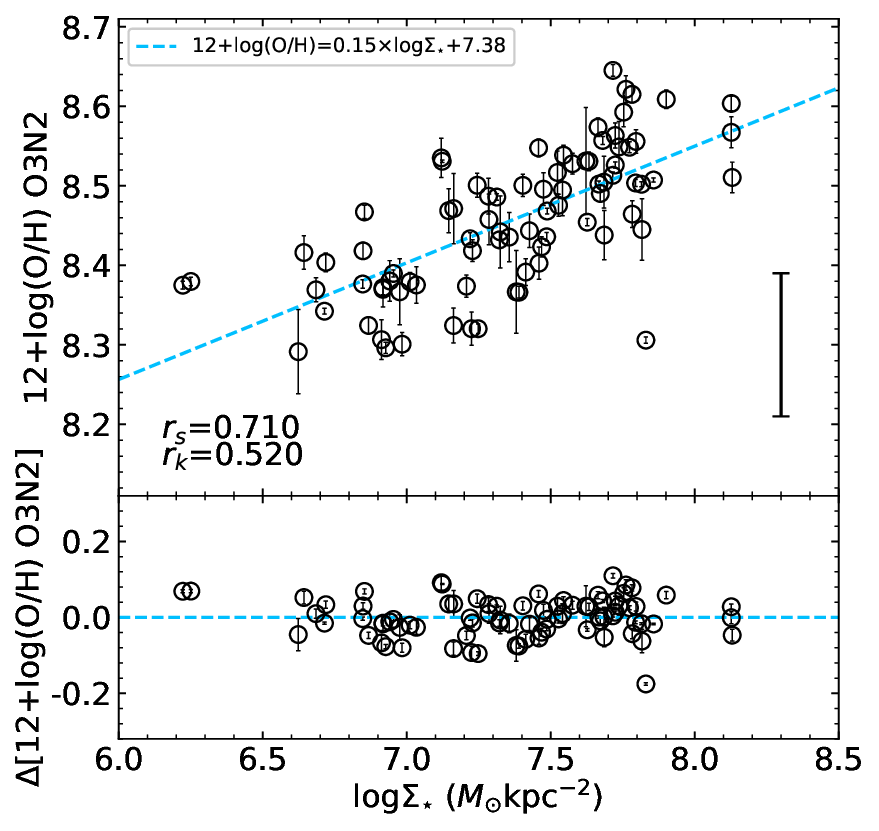}}}
\caption{Spatially resolved stellar surface mass density - gas metallicity relation for NGC 628. Upper: the blue dashed lines indicate the linear fit to the data. The vertical error bar represents the systematic uncertainty of the O3N2 calibrator from \citet{marino13}. Bottom: the residual $\Delta$12+log(O/H) as a function of log${\Sigma}_{\star}$.}
\label{fig:mzr}
\end{figure}

In Figure \ref{fig:mzr}, the rMZR of NGC 628 is presented, using the derived the oxygen abundance from spectra and the derived ${\Sigma}_{\star}$ from the SED fitting with multi-wavelength photometric data. This plot illustrates a clear correlation between ${\Sigma}_{\star}$ and gas-phase abundance at local scales of  $\sim$100 pc. To elucidate the shape of the rMZR, we employ an asymptotic function proposed by \citet{moustakas11}, represented as $y = a + b(x - c)e^{-(x-c)}$, to characterize the relationship between ${\Sigma}_{\star}$ and 12+log(O/H). The threshold point at ${\Sigma}_{\star}$ $\sim10^{9}\rm{M_{\odot}kpc^{-2}}$ mark the shift from  the linear trend at lower ${\Sigma}_{\star}$ values to the plateau observed at higher ${\Sigma}_{\star}$ levels \citep{erroz19}. In our analysis, the rMZR is determined only up to ${\Sigma}_{\star}$ of $\sim10^{8.2}\rm{M_{\odot}kpc^{-2}}$ within the sample. Consequently, a linear regression model is applied to establish the relationship for NGC 628. The slope of the rMZR is calculated as $0.15\pm0.02$ dex/log${\Sigma}_{\star}$, which closely aligns with the slope of $0.12\pm0.01$ derived from double-linear function for the region below  $\sim10^{9}\rm{M_{\odot}kpc^{-2}}$ in the study by \citet{erroz19}. 

\begin{figure*}[!htb]
\center
\resizebox{\width}{!}{
{\includegraphics[width=0.9\textwidth]{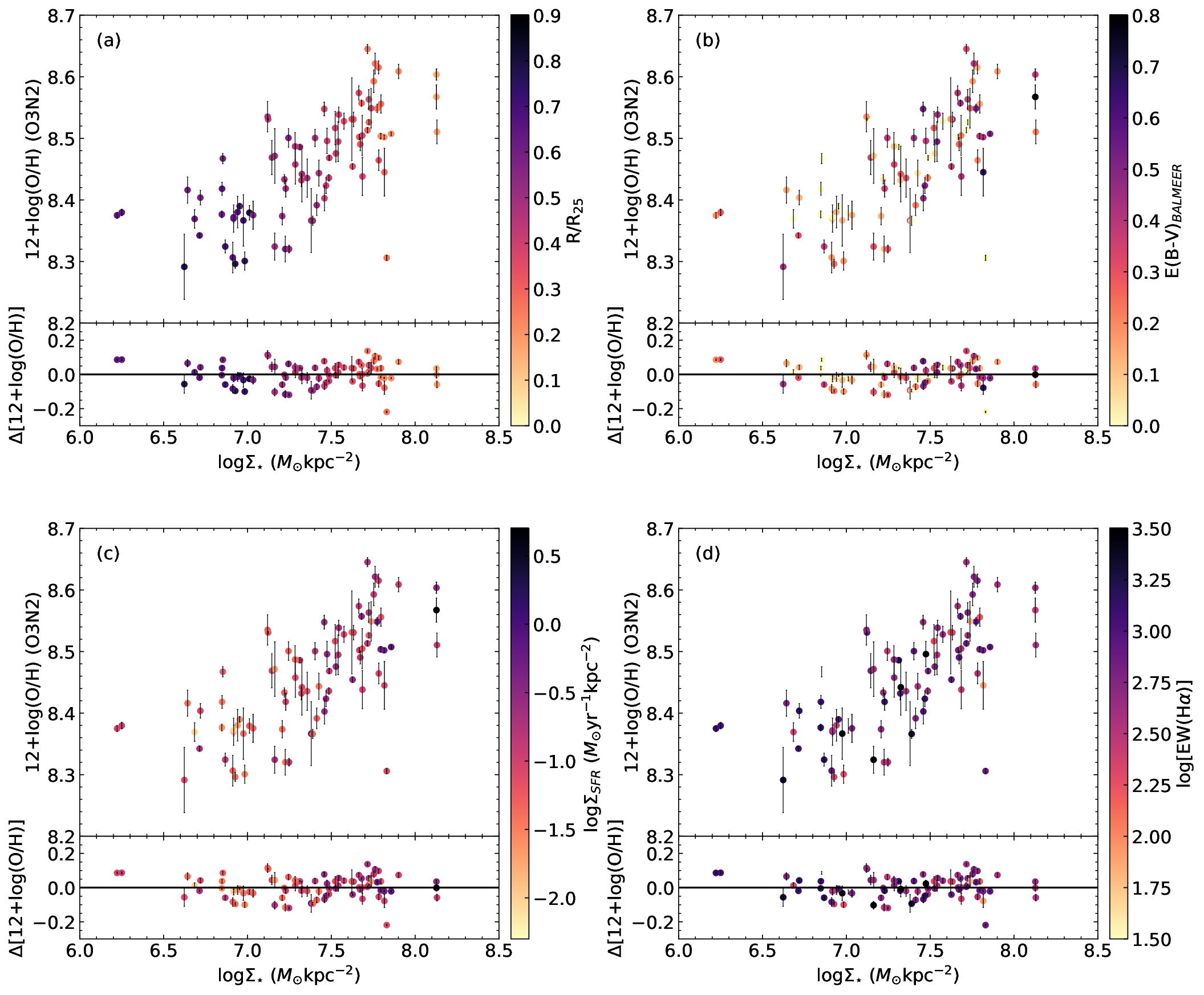}}}
\caption{The rMZR color-coded with various physical parameters including galactocentric distance (a), dust extinction (b),  $\Sigma_{\mathrm{SFR}}$ (c) and EW({\Ha}) (d). The dependence of rMZR on these four parameters are presented.}
\label{fig:mzr_four}
\end{figure*}

In Figure \ref{fig:mzr_four}, we investigate the dependence of rMZR on various galaxy properties. The top-left panel illustrates the galactocentric distance varying with ${\Sigma}_{\star}$ and 12+log(O/H). As discussed in Section \ref{subsubsec:metallicity}, both ${\Sigma}_{\star}$ and gas-phase abundance decrease as galactocentric distance increases, supporting an ``inside-out'' scenario in NGC 628. The central regions in NGC 628 form earlier and are denser, leading to more chemically evolved than less-dense regions. From the top-right panel of Figure \ref{fig:mzr_four}, we observe that dust extinction increases with gas-phase abundance and stellar mass. However, further analysis of residuals indicates that stellar mass play a more significant role in this relationship. The bottom-left panel displays the relationship between the $\Sigma_{\mathrm{SFR}}$, ${\Sigma}_{\star}$ and 12+log(O/H) in NGC 628. It is anticipated that $\Sigma_{\mathrm{SFR}}$ rises with ${\Sigma}_{\star}$ following the resolved star formation main sequence discussed in Section \ref{subsubsec:rsfms}. Additionally, lower $\Sigma_{\mathrm{SFR}}$ values are observed in regions with lower gas-phase abundance, while a wide range of $\Sigma_{\mathrm{SFR}}$ values is distributed within $8.3 < $ 12+log(O/H) $< 8.6$. It remains unclear whether the relationship between rMZR and $\Sigma_{\mathrm{SFR}}$ arises from the rSFMS or is influenced by SFR as a secondary parameter within rMZR. The bottom-right panel of Figure \ref{fig:mzr_four} demonstrates that EW({\Ha}) is not strongly correlated to rMZR.

\subsubsection{rMZR dependent on {\HI} mass surface density}\label{subsubsec:himzr}

\begin{figure*}[!htb]
\center
\resizebox{\width}{!}{
{\includegraphics[width=1.0\textwidth]{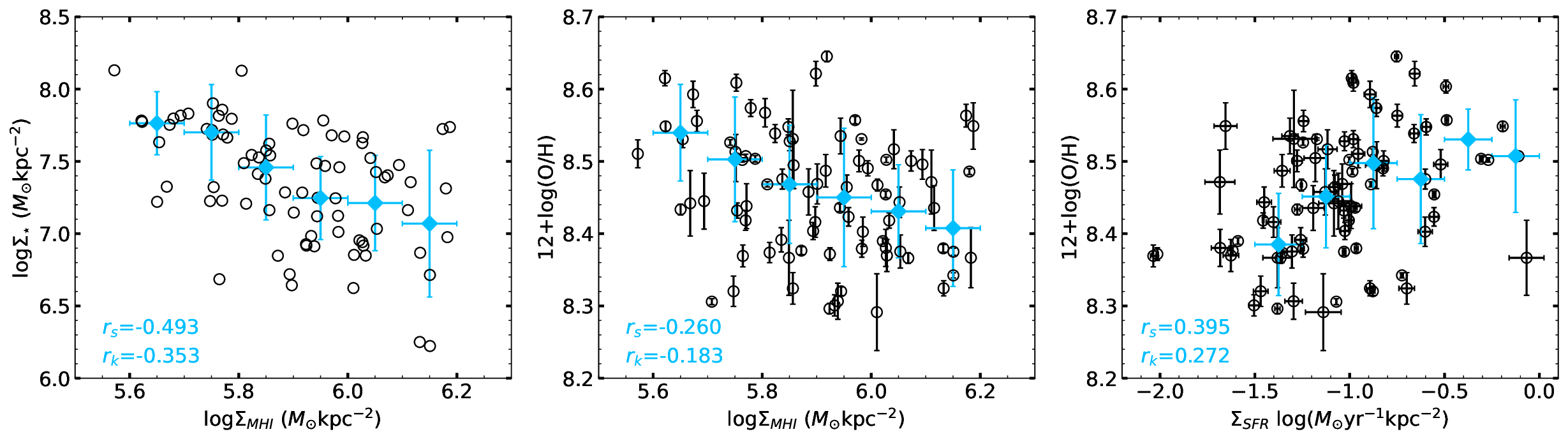}}}
\caption{The dependence of ${\Sigma}_{\HI}$ and $Y$. From left to right, $Y$ is defined as ${\Sigma}_{\star}$, gas-phase oxygen abundance and $\Sigma_{\mathrm{SFR}}$, respectively. In each panel, the blue diamonds represent the median values for each radial bin, with accompanying vertical error bars indicating the median and its standard deviation. The horizontal bars denotes the range of each bin.}
\label{fig:himzr}
\end{figure*}

\begin{figure}[!htb]
\center
\resizebox{\width}{!}{
{\includegraphics[width=0.45\textwidth]{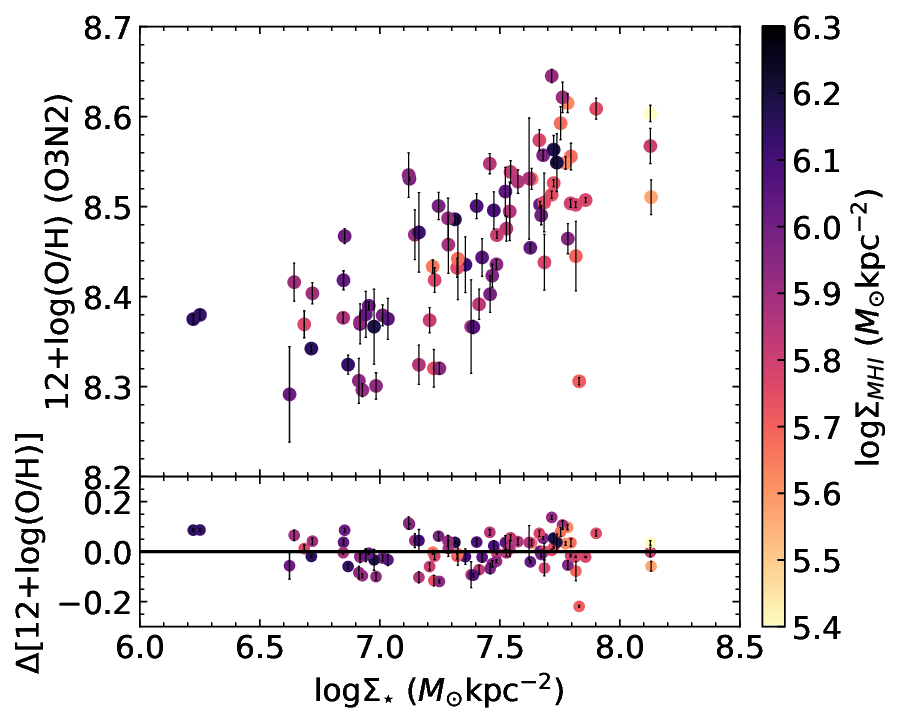}}}
\caption{rMZR color-coded with ${\Sigma}_{\HI}$.}
\label{fig:mzrhi}
\end{figure}

In Section \ref{subsubsec:rmzr}, the relationship between the rMZR and $\Sigma_{\mathrm{SFR}}$ in NGC 628 appear unclear. This ambiguity arises from the fact that the gas influencing metal content also contributes to star formation within the framework of $\Lambda$ cold dark matter cosmology. Recent observations have suggested a secondary dependence of the MZR on molecular gas mass and {\HI} mass \citep{bothwell16,brown18,zu20}. Notably, \citet{chen22} have found that the inner {\HI} mass within the optical radius can marginally reduce the scatter in the MZR relation based on {\HI} follow-up observations from the SDSS-IV MaNGA survey. Utilizing spatially resolved {\HI} imaging of NGC 628 at a resolution of 6\arcsec, obtained as part of the THINGS sample, we investigate the relationship between rMZR and ${\Sigma}_{\HI}$. The coefficients $r_{s}$ and $r_{k}$ shown in Figure \ref{fig:himzr} indicate an inverse correlation between gas-phase oxygen abundance and ${\Sigma}_{\HI}$, while a positive correlated is observed with $\Sigma_{\mathrm{SFR}}$. In regions with high ${\Sigma}_{\HI}$ generally exhibit lower metallicity, likely due to {\HI} gas representing a more ``pristine" reservoir that has undergone minimal stellar processing. This inflowing {\HI} gas, particularly prominent in the outskirts of the galaxy, may dilute local metallicity, thereby modulating the rMZR in specific regions. Figure \ref{fig:mzrhi} demonstrates that ${\Sigma}_{\HI}$ does not fundamentally alter the rMZR's primary dependence on ${\Sigma}_{\star}$. Thus, the influence of ${\Sigma}_{\HI}$ is largely restricted to a localized modulation effect rather than serving as a secondary dependency in the rMZR. This analysis indicates that ${\Sigma}_{\HI}$ plays a minor, supportive role in shaping rMZR, especially through dilution effects in the outer galaxy, yet it does not significantly impact the overarching rMZR trend predominantly driven by ${\Sigma}_{\star}$.

As discussed in Section \ref{subsubsec:azimuthal}, a significant portion of {\HI} is contained within the extended outer disk of NGC 628 and gradually moves towards the center. By examining the relationships between ${\Sigma}_{\HI}$-${\Sigma}_{\star}$ and ${\Sigma}_{\HI}$-$Z_{gas}$, it can be deduced that post-gas accretion, star formation may increase due to the hightened  stellar mass and gas abundance in the disk and ultimately leading to the enrichment of metals. Essentially, the gas extended outer disk can be refilled into areas where star are forming. Consequently, NGC 628 exemplifies a typical ``inside-out" disk galaxy driven by gas accretion.

\subsubsection{Relationship between ${\Sigma}_{\star}$ and $\Sigma_{\mathrm{SFR}}$
} \label{subsubsec:rsfms}
\begin{figure}[!htb]
\center
\resizebox{\width}{!}{
{\includegraphics[width=0.45\textwidth]{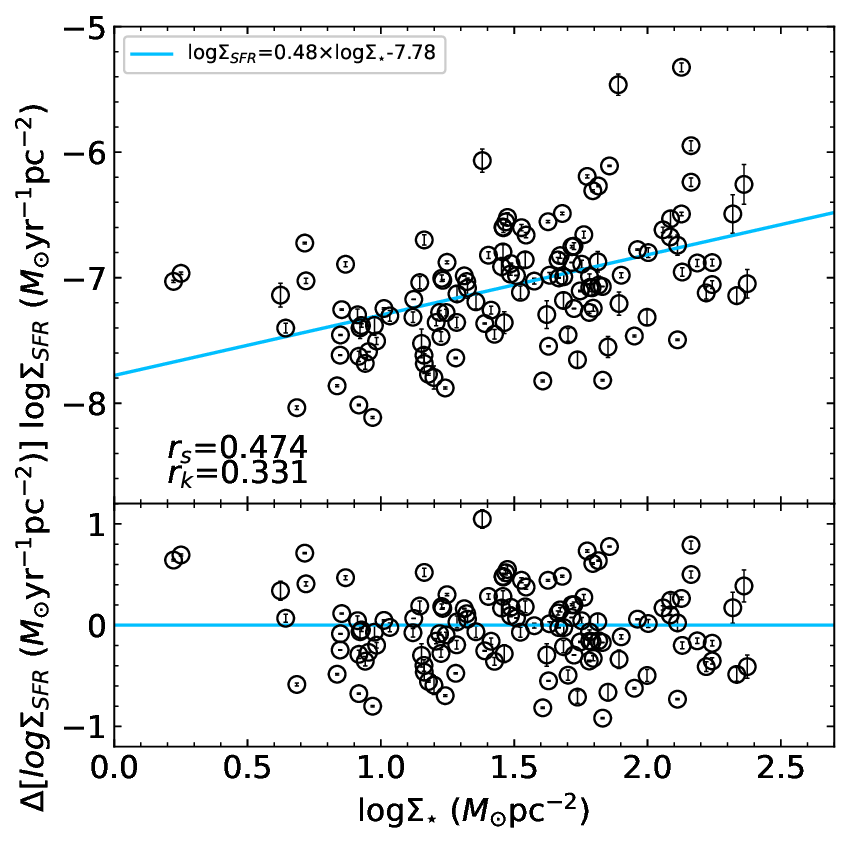}}}
\caption{Spatially resolved stellar surface mass density - SFR surface density relation for NGC 628. Upper: the blue solid line presents the linear fit to the data. Bottom: the residuals $\Delta$log$\Sigma_{\mathrm{SFR}}$ as a function of log${\Sigma}_{\star}$.}
\label{fig:rsfms}
\end{figure}

To investigate the star formation process at a local level in NGC 628, we present the linear correlation between $\Sigma_{\mathrm{SFR}}$ and ${\Sigma}_{\star}$ in Figure \ref{fig:rsfms}. This relationship spans 2.5 orders of magnitude in ${\Sigma}_{\star}$ and three orders in $\Sigma_{\mathrm{SFR}}$. We observe a positive correlation between $\Sigma_{\mathrm{SFR}}$ and ${\Sigma}_{\star}$, although some scatter is present. A linear fit yields a slope of 0.48$\pm$0.08 dex/log${\Sigma}_{\star}$ and an intercept of -7.61$\pm$0.13 dex/log$\Sigma_{\mathrm{SFR}}$. In comparison, \citet{erroz19} reported a slope of 0.79$\pm$0.07 dex/log${\Sigma}_{\star}$ and an intercept of -9.54$\pm$0.20 dex/log$\Sigma_{\mathrm{SFR}}$, based on MUSE data for star-forming regions at $\sim$100 pc scales. This analysis utilized a sample with a ${\Sigma}_{\star}$ spanning five orders of magnitude. Their results align closely with those obtained from CALIFA \citep{cano16} and MaNGA \citep{hsieh17} data for star-forming regions at kpc scales. \citet{erroz19} additionally applied a double-line fit to the rSFMS at a threshold ${\Sigma}_{\star}$ of \(10^{3} \, M_{\odot} \, \text{pc}^{-2}\). For ${\Sigma}_{\star}$ values below \(10^{3} \, M_{\odot} \, \text{pc}^{-2}\), they obtained a fitted slope of \(0.37 \pm 0.04\), closely matching the slope observed in NGC 628 within this density range.

\begin{figure*}[!htb]
\center
\resizebox{\width}{!}{
{\includegraphics[width=0.9\textwidth]{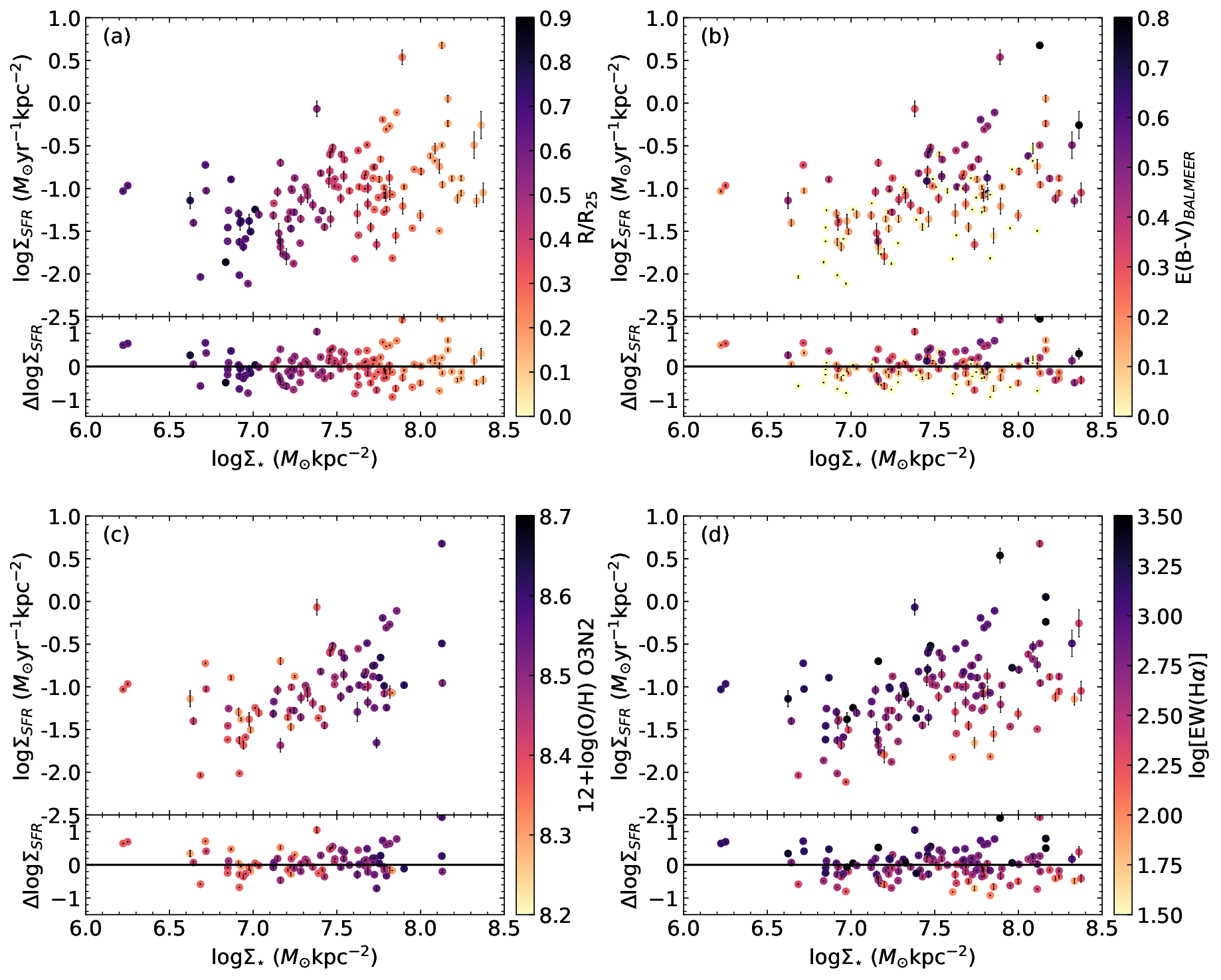}}}
\caption{The rSFMS color-coded with different physical parameters of galactocentric distance (a), dust extinction (b), 12+log(O/H) (c) and EW({\Ha}) (d). The dependence of rSFMS on these four parameters are presented.}
\label{fig:rsfms_four}
\end{figure*}

Just as we investigate the dependence of rMZR with galaxy properties, the third parameters are color-coded in the ${\Sigma}_{\star}$-$\Sigma_{\mathrm{SFR}}$ diagram in Figure \ref{fig:rsfms_four}. As described in section \ref{subsubsec:esfr}, $\Sigma_{\mathrm{SFR}}$ increases with galactocentric distance, but presents a relatively large dispersion in the top-left panel. The top-right panel presents that the extinction increases with increasing $\Sigma_{\mathrm{SFR}}$ but does not depend on ${\Sigma}_{\star}$ indicating an association between dust production and the emergence of new stars. Notably, this increase in extinction appears decoupled from the influence of stellar mass and metallicity, suggesting that dust generation is primarily tied to the process of star formation itself, rather than being directly dependent on pre-existing stellar mass or metal content. The bottom-left panel shows the dependence of gas-phase abundance with $\Sigma_{\mathrm{SFR}}$ and ${\Sigma}_{\star}$. As ${\Sigma}_{\star}$ increases, a corresponding rise in $\Sigma_{\mathrm{SFR}}$ is observed, along with an increase in 12+log(O/H). Figure \ref{fig:himzr} further demonstrates that 12+log(O/H) tends to increase with $\Sigma_{\mathrm{SFR}}$, albeit with notable scatter. This indicates a complex interdependence among these three parameters, making it insufficient to assert that 12+log(O/H) serves as a secondary dependent parameter within the rSFMS relationship. The bottom-right panel shows the variation of equivalent width of the {\Ha} emission line with ${\Sigma}_{\star}$ and $\Sigma_{\mathrm{SFR}}$ in NGC 628. From the residuals $\Delta$log$\Sigma_{\mathrm{SFR}}$ as a function of log${\Sigma}_{\star}$, higher $\Sigma_{\mathrm{SFR}}$ has a higher EW({\Ha}).

\section{Summary} \label{sec:summary}
NGC 628 serves an excellent laboratory for investigating the spatial distributions of physical properties due to its low inclination and proximity. Using multi-band photometric data spanning ultraviolet to infrared wavelengths, we have derived spatially resolved maps of age, metallicity, and reddening for NGC 628, as presented in \citet{zou2011c}. In this study, we observed a significant number of {\HII} regions in NGC 628 utilizing the the long-slit spectrograph of the NAOC 2.16 m telescopes. Leveraging these spectral and multi-wavelength photometric data, we determined various physical properties, including gas-phase extinction, SFR surface density, oxygen abundance, and stellar mass surface density. We examined the two-dimensional distributions and corresponding radial gradients of these properties to investigate the evolutionary clues. Additionally, we explored the rMZR and rSFMS at a resolution of $\sim$100 pc, along with their potential dependencies on third  parameters. are also probed. The main conclusions are summarized as follows:
\begin{enumerate}[(i)]
\item The azimuthal distributions of oxygen abundance, dust extinction, EW({\Ha}), and $\Sigma_{\mathrm{SFR}}$ in NGC 628 are nearly uniform, indicating that it is a relatively isolated galaxy with minimal interactions with neighboring galaxies. Analysis of the FAST {\HI} morphology and kinematics further supports this conclusion, suggesting that NGC 628 has not experienced recent interactions in its evolutionary history.
\item A mild radial extinction gradient is present in NGC 628, although it exhibits a relatively large dispersion. The local physical conditions predominantly influence the dust properties within this galaxy. Consistent with most spiral galaxies, NGC 628 shows a positive gradient in EW({\Ha}), highlighting that the inner disk has undergone a longer evolutionary process. The inner disk also displays higher $\Sigma_{\mathrm{SFR}}$ compared to the outer disk. NGC 628 features a radial negative gradient in gas-phase metallicity across its disk (-0.443 dex $R_{25}^{-1}$). The correlated positive gradient in EW({\Ha}), decrease in radial $\Sigma_{\mathrm{SFR}}$ and the negative gradient in gas-phase metallicity support the ``inside-out" galaxy growth scenario in NGC 628.
\item The correlation between gas-phase metallicity and $\Sigma_{\star}$ in NGC 628 exhibits a linear relationship with a slope of 0.14 dex/log$\Sigma_{\star}$ within 0.9 $R/R_{25}$ at local scales of $\sim$100 pc. This slope is consistent with the rMZR observed in the MUSE Atlas of Disks survey, which also operates at a spatial resolutions of around 100 pc. We found no significant secondary dependencies of the rMZR on $E(B-V)$, $\Sigma_{\mathrm{SFR}}$, EW({\Ha}), or ${\Sigma}_{\HI}$. Additionally, the rSFMS holds for {\HII} regions in NGC 628, with a slope of 0.48 dex/log$\Sigma_{\star}$. There is also a trend of increasing dust extinction and EW({\Ha}) with higher $\Sigma_{\mathrm{SFR}}$.
\item {\HI} kinematic features from FAST reveal that NGC 628 currently undergoing gas accretion from its extended outer disk into the interior. Analysis of the relationships between ${\Sigma}_{\HI}$ and gas-phase metallicity, as well as between ${\Sigma}_{\HI}$ and $\Sigma_{\star}$, indicates that the accreted gas can inhabit star-forming regions within the galaxy, thereby fueling ongoing star formation. This scenario exemplifies gas accretion facilitating "inside-out" growth in disk galaxies.
\end{enumerate}

So far, as predicted by the gas regulator models, some observations have revealed that the MZR is more strongly linked to molecular and neutral gas mass \citep{bothwell16,chen22}, with a weaker dependence on the SFR, implying that SFR may indirectly reflect the influence of neutral gas on metallicity. Our analysis of NGC 628 finds no clear evidence that rMZR is affected by ${\Sigma}_{\HI}$. We plan to leverage upcoming Dark Energy Spectroscopic Instrument (DESI) data and spectra of {\HII} regions in nearby galaxies from NAOC 2.16m telescope, combined with high-resolution synthesis data from THINGS and single-dish from FAST, to further explore the relationship between ${\Sigma}_{\HI}$ and rMZR.

\acknowledgments
We thank the anonymous referee for constructive comments. This work is supported by Chinese Academy of Sciences (CAS) ``Light of West China" Program (No. 2021-XBQNXZ-029) and the National Key R\&D Program of China (grant Nos. 2022YFA1602902 and 2023YFA1607804), the National Natural Science Foundation of China (grant No. 12120101003), the Natural Science Foundation of Xinjiang Uygur Autonomous Region (No. 2020D01B59), the National Key R\&D Program of China for Intergovernmental Scientific and Technological Innovation Cooperation Project (No. 2022YFE0126200). This work is also supported by the National Key R\&D Program of China (grant Nos. 2023YFA1607800,  2023YFA1608100, and 2023YFF0714800), the National Natural Science Foundation of China (NSFC; grant Nos. 12373010, 12173051, and 12233008), and the Beijing Municipal Natural Science Foundation (grant No. 1222028), the science research grants from the China Manned Space Project with Nos. CMS-CSST-2021-A02 and CMS-CSST-2021-A04 and the Strategic Priority Research Program of the Chinese Academy of Sciences with Grant Nos. XDB0550100 and XDB0550000. The authors acknowledge the science research grants from Tianshan Talent Training Program (No. 2023TSYCLJ0053). 

This work uses the observational time of the 2.16m telescope at the Xinglong station of the National Astronomical Observatories of China and the observational time of the MMT telescope obtained via the Telescope Access Program (TAP), which is funded by the National Astronomical Observatories of China, the Chinese Academy of Sciences (the Strategic Priority Research Program, ``The Emergence of Cosmological Structures” grant No. XDB09000000), and the Special Fund for Astronomy from the Ministry of Finance. The 2.16m telescope is jointly operated and administrated by the National Astronomical Observatories of China and Center for Astronomical Mega-Science, CAS.

\end{CJK*}
\end{document}